\documentclass[11pt]{amsart}

\usepackage{hyperref}

\usepackage{amsmath, amsthm, amssymb, delarray, todonotes, graphicx, subcaption, float, url}

 \usepackage[margin=1in]{geometry}
\usepackage{enumitem}
\setlist[enumerate, 1]{label=\arabic*.}
\setlist[enumerate, 2]{label=\alph*.}

\newcommand{\R}{{\mathbb R}}

\begin{document}


\title[TDA and COVID-19]{Topological data analysis model for the spread of the coronavirus}

\author{Yiran Chen}
\address{Department of Mathematics, Wellesley College, 106 Central Street, Wellesley, MA 02481}
\email{ychen10@wellesley.edu}
\urladdr{}

\author{Ismar Voli\'c}
\address{Department of Mathematics, Wellesley College, 106 Central Street, Wellesley, MA 02481}
\email{ivolic@wellesley.edu}
\urladdr{ivolic.wellesley.edu}



\begin{abstract}
We apply topological data analysis, specifically the Mapper algorithm, to the U.S.
COVID-19 data.  The resulting Mapper graphs provide visualizations of the pandemic
that are more complete than those supplied by other, more standard methods. They
encode a variety of geometric features of the data cloud created from geographic
information, time progression, and the number of COVID-19 cases.  They reflect the
development of the pandemic across all of the U.S.~and capture the growth rates as
well as the regional prominence of hot-spots. The Mapper graphs allow for easy
comparisons across time and space and have the potential of becoming a useful
predictive tool for the spread of the coronavirus.
\end{abstract}


\maketitle

\tableofcontents

\parskip=6pt
\parindent=0cm



\section{Introduction}


Topological data analysis (TDA) is a method for understanding data clouds that
attempts to gain insight into the data by treating it as a geometric object and
extracting information based on its ``shape''. There are several TDA instruments
available, and the one we use in this paper is called the \emph{Mapper
  algorithm}. Introduced by Singh, M\' emoli, and Carlsson~\cite{SMC:Mapper}, Mapper
is a way to simplify data while preserving many of its topological features. The idea
is to project complicated data in a way that makes the projection tractable, then use
topology to cover the projection with certain sets, and then look at the preimages of
those sets under the projection map. This information is then used to construct a
graph that retains many of the topological information about the original data set
such as clustering, connectedness, 1-dimensional holes, etc. This graph lends itself
more readily to analysis than the original data set, but is still complicated enough
to provide more insight than some other dimension-reduction and clustering techniques
such as principal component analysis. Mapper has been used in a variety of situations
such as breast cancer data~\cite{NLC:BreastCancer}, image
processing~\cite{RHR:TDAImage}, and patent
development~\cite{EHIO:MapperFirms}.\footnote{Mapper was even used to analyze NBA
  player characteristics; see for example
  \href{https://www.wired.com/2012/04/analytics-basketball/}{this article}.} For
overviews on Mapper and TDA more generally, see~\cite{Carlsson:TDA,
  Carlsson:TDAHomotopy}.

In this paper, we apply the Mapper algorithm to the United States COVID-19 data. We
consider the 4-dimensional data cloud consisting of the longitudes and latitudes of
the U.S.~counties and territories, number of days elapsed in the pandemic (starting
on January 22, 2020), and the number of cumulative COVID-19 cases recorded by that
day in each county.

The resulting Mapper graphs are rich in information and more complete than other more
standard methods of visualization. They reflect important aspects about the
development and spread of the COVID-19 pandemic across the U.S. and capture dramatic
growth rates, trends, and regional prominence of hot-spots. In particular, compared
with traditional ways of data visualization that similarly provide real-time
monitoring of the spread of COVID-19 across the
U.S.~(e.g. \href{https://www.arcgis.com/apps/opsdashboard/index.html#/bda7594740fd40299423467b48e9ecf6}{Johns
  Hopkins University's COVID-19 dashboard}), Mapper visualization has the following
advantages:
\begin{itemize}
\item It captures the entire developmental process of the outbreak in the U.S.,
  including the emergence of hot-spots across time. Instead of displaying a plain
  snapshot of all the COVID-19 cases in the U.S.~at certain point in time or a time
  series data for a single location, a Mapper representation captures the evolution
  of the spread in the entire country.

\vspace{4pt}

\item The Mapper graph makes it easy to compare data across time and space. A certain
  U.S.~county's position in the graph is in part determined relative to the
  surrounding counties so, when irregular patterns appear in the visualization, it is
  because the data it represents has some volatility in comparison with its
  neighboring counties. This makes it easier to spot regional hot-spots while
  retaining an overarching view of the United States.
\end{itemize}

In Section~\ref{ssec:algorithm}, we give background on the Mapper algorithm, trying
to keep the technical aspects to the minimum for clarity and
readability. Sections~\ref{ssec:data} and~\ref{ssec:COVIDMapper} explain how we apply
the Mapper algorithm to the U.S.~COVID-19 data.

Section~\ref{sec:results} contains the analysis of the various Mapper images we
obtain. We discuss the meaning of connected components in
Section~\ref{ssec:components} and explain how each of the coordinates -- time,
geography, and the number of cases -- influence the connectedness of the graph. We
turn to the other most prominent feature of the Mapper, its branches, in
Section~\ref{ssec:branches}, and discuss how these structures indicate the appearance
and development of COVID-19 hot-spots. Finally in Section~\ref{sec:evolution}, we take
a look at the evolution of the Mapper with respect to time and examine how the
succession of graphs provides useful information about the development of the
pandemic across time and space.

In Section~\ref{sec:future}, we make some remarks about the many ways in which the
work in this paper could be extended and generalized. For example, the Mapper could
be overlaid with dates of stay-at-home orders so that the effectiveness and
timeliness of such directives could be assessed. Socio-economic factors could also be
read into the Mapper so that the COVID-19 spread could be correlated with such data.
In addition, our analysis can be replicated for any region in the world, and, for
some of them, travel restrictions between countries and border closings could also be
incorporated into the data to gauge their efficacy. In addition, the branches that
indicate hot-spots could be given more rigorous graph-theoretic structure that could
lead to a robust predictive model. Finally, another TDA tool called \emph{persistent
homology} could be employed and used in conjunction with Mapper to gain an even
deeper understanding of the spread of the COVID-19 pandemic.

Due to its relatively recent emergence, topological data analysis has not yet been
applied extensively to epidemiology.  The idea that this methodology could be useful
in the study of viral evolution was put forth in~\cite{CCR:TDAViral}. The paper
\cite{TKHKMPM} provides a general framework for using TDA to study contagion across
networks. Applications to particular diseases can be found in
\cite{CS:TDAEpidemiology}, which studies the spread of influenza, and
\cite{LP:TDAZika}, which does the same for Zika. All of the above papers use the
persistent homology arm of TDA. The Mapper algorithm does not appear to have been
used in the context of epidemiology yet, except in~\cite{DR:COVIDBallMapper}, which
uses the \emph{Ball Mapper} to examine economic and COVID-19 case data in England;
this also seems to be the only article that studies the coronavirus pandemic through
TDA to date.

\subsection{Acknowledgements} The second author would like to thank the Simons
Foundation for its support.


\section{Methods}\label{sec:methods}


\subsection{The Mapper algorithm}\label{ssec:algorithm}


Given a potentially high-dimensional data set $X$ (each data point is possibly a
vector in some high-dimensional Euclidean space), Mapper replaces it by a graph
consisting of nodes and edges that retain information about some of the original
geometric features of the data (more generally, Mapper creates a higher-dimensional
version of a graph called a \emph{simplicial complex}, but we will stick to graphs in
this paper).  Not only are various important features such as components and holes
preserved, but Mapper's visual representation of the data can often reveal new
structures and insights like flares or branches that other, more traditional
statistical methods, cannot.

To implement Mapper:
\begin{enumerate}
  \item Use some projection function (also called \emph{filter} or \emph{lens})
$f\colon X\to \R^d$ to map the dataset into some Euclidean space $\R^d$.

  \item Cover $f(X)$, the image of the dataset in $\R^d$, by a collection of
overlapping open sets $$\mathcal U=\{U_i\}_{i\in I}.$$ Here $I$ is some finite
indexing set. Sets $U_i$ are also sometimes called \emph{bins}.

  \item Apply some clustering algorithm to each preimage $f^{-1}(U_i)$ (preimages are also
called \emph{fibers}). This produces $i_j$ clusters $C_{i_1}, C_{i_2}, \ldots, C_{i_j}$
in the $i$th preimage.

  \item  Create a graph whose vertices or nodes consist of the set of all clusters
\[\{C_{i_1}, C_{i_2},\ldots, C_{i_j},\ 1\leq i\leq |I|\}.\]

An (unoriented) edge between nodes $C_{j_i}$ and $C_{k_l}$ is added if and only if
\[
  C_{j_i}\cap C_{k_{l}}\neq \emptyset.
\]
\end{enumerate}
This collection of nodes and edges is the Mapper graph, denoted by $M(X,\mathcal
U,f)$.\footnote{If one were to create the more complicated simplicial complex that
carries even more information than the Mapper graph, then one would also look at all
$n$-fold intersections for $n>1$ to determine the $n$-simplices of the complex.}

\begin{figure}[h]
\includegraphics[width=0.8\linewidth]{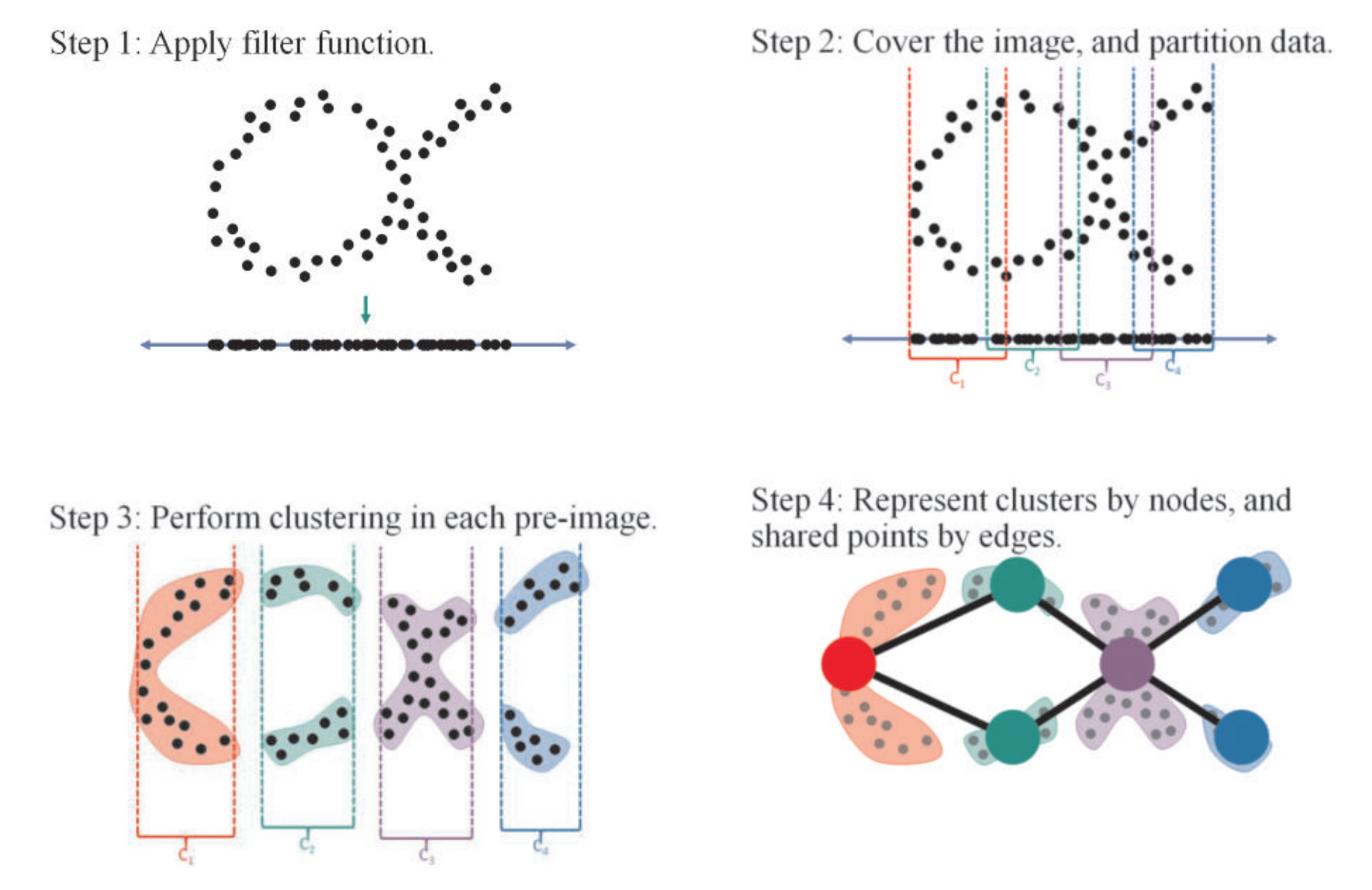}
\caption{An illustration of the Mapper algorithm. Image Source: Escolar, Hiraoka,
  Igami, Ozcan~\cite{EHIO:MapperFirms}.}\label{fig:Mapper_example}
\end{figure}

Figure~\ref{fig:Mapper_example} illustrates this procedure on a simple dataset $X$
that lives in $\R^2$, with the projection function $f\colon X\to \R$ mapping to the
real line by forgetting the first coordinate.

The projection function depends on the context and on the aspect of the data that is
deemed most relevant. It could simply forget coordinates, but most useful projections
are statistically more meaningful, such as the kernel density estimator, distance to
measure, or graph Laplacian. In the applications of Mapper to the study of cancer,
for example, a Healthy State Model which decomposes tumor cells into the ``disease''
and ``healthy'' components is used to project to the disease
part~\cite{NLC:BreastCancer}. The domain of the projection is often $\R$, i.e.~the
projection simply assigns a real number to each data vector.

The image $f(X)$ is usually covered by hypercubes (products of intervals), but open
covers of any shape are a priori allowed. The number and the amount of overlap of the
open sets determines many features of the graph, including the number of its
connected components.

The task of a clustering algorithm is to determine which subsets of data points in
each preimage appear to be closer to each other than to others.  This is essentially
the discrete version of identifying the connected components of a topological space.
Clustering can be done by selecting an appropriate metric or distance function
(Euclidean, correlation, etc.) on the data and using it to decide the ``closeness''
via a procedure such as \emph{single-linkage hierarchical clustering}. In fact, there
is no need for the distance function to satisfy the triangle inequality, so
technically only a \emph{semimetric} is required. Clustering can also be done in the
image $f(X)$, but some information loss due to projection is possible.

The size of the nodes can also be regulated and it corresponds to the number of data
points in that cluster. Nodes can also be colored according to some chosen
characteristics of the data. For example, in the applications of Mapper to breast
cancer data~\cite{NLC:BreastCancer}, the nodes are colored according to survival
rates.

Note that there are many choices that are left to the user -- the projection
function, the number of open sets in the cover of $f(X)$ and the amount of their
overlap, and the clustering algorithm. This ad hoc nature of the algorithm is in many
ways its advantage because it allows for great generality and flexibility in the
types of data that can be treated and the questions that can be asked about it.

Mapper is an unsupervised data analysis method, but is more subtle than other ones
such as principal component analysis. It performs dimensionality reduction and
clustering while preserving the shape of data. This is possible because Mapper
clusters locally but then extrapolates back to global data via the edges. The graph
is often a lot easier to interpret than other visualizations like scatterplots and it
effectively captures many facets of the original, high-dimensional data.

There are various free technical implementations of the Mapper algorithm, such as
Python Mapper~\cite{PythonMapper}, TDAmapper (R package)~\cite{TDAmapper}, and Kepler
Mapper (which we use in this paper)~\cite{KeplerMapper2019}.

\subsection{COVID-19 data}\label{ssec:data}%


The data we use in this study comes from the COVID-19 Data Repository by the Center
for Systems Science and Engineering (CSSE) at Johns Hopkins
University.\footnote{There are unclassified data labeled as ``Out of state'' or
``Unassigned'' in the original data set provided by CSSE. In order to preserve the
coherency and usefulness of our analysis, we do not incorporate such data into our
point cloud.} Specifically, we use a collection of time series data on the number of
confirmed COVID-19 cases reported by 3155 U.S.~counties (and territories) starting on
1/22/20. The end date for most of our analysis will be 6/19/20, but some of it will
go up to 7/24/20 (see Figures~\ref{fig:update-all} and~\ref{fig:update-filtered} in
particular). Each data point encodes information on the geographic location of a
certain U.S.~county, a date, and the number of cumulative confirmed COVID-19 cases
reported in that county on that date.

More precisely, suppose a U.S. county is located at latitude $x$ and longitude
$y$. Here $x$ is positive when it represents a northern latitude and negative when it
represents a southern one (e.g.~American Samoa). Similarly $y$ is positive if it is
an eastern longitude (e.g.~Guam) and negative otherwise.

Let the coordinate $w$ encode the date information as relative to the first date in
our time series, namely 1/22/20, which is itself given value 0. Lastly, let $z$ be
the number of cumulative confirmed COVID-19 cases in that county on date $w$. Then
our cloud consists of data points
\[p = (x, y, z, w) = (\text{latitude}, \text{longitude}, \text{cumulative confirmed
    cases}, \text{day}).
\]

For example, King County in Washington state is located at roughly $47^{\circ} N,
122^{\circ} W$ and reported 7472 cumulative confirmed COVID-19 cases on 5/17/20. In
our original point cloud, this data point is thus encoded as
\[(47.49137892, -121.8346131, 7472, 117).\]

The entire data cloud up to 5/17/20 will then contain
\[\text{(number of counties)}\cdot\text{(number of days)} = 3155\cdot 117 = 369,135\]

vectors of length 4, i.e.~elements of $\R^4$.


\subsubsection{Normalization}\label{sssec:normalization}%


Since the four coordinates of our data vector use varying systems of measurement,
which cause their numerical values to have different orders of magnitude, modeling
them directly into a 4-dimensional Euclidean space to serve as the basis for Mapper
clustering means that the four coordinates will take on disproportionate weights in
determining the shape of the Mapper graph. More specifically, coordinates that have
relatively small numerical values, such as the time coordinate, will barely be
evident in the Mapper graph. We hence need to offset the distortion effect induced by
the numerical discrepancies in the coordinates of our data point. We achieve this by
separately scaling each coordinate of the input row vectors to column-wise unit norm.
This means that if we square a chosen coordinate of all the scaled vectors and then
sum them within its feature, the total would be 1.

From now on, we will be working with this normalized point cloud. Note that, as the
time interval is modified (by changing the end date), the values of the coordinates
of the vectors change under the normalization as well, i.e.~the same vector might be
normalized to different values after the time frame is extended to a later date.


\subsection{Application of the Mapper algorithm to the COVID-19 data}\label{ssec:COVIDMapper}


Recall the general description of the Mapper algorithm from
Section~\ref{ssec:algorithm}. To construct a Mapper graph for our point cloud $X$, we
use the Python implementation of the Mapper, KeplerMapper by Van Veen and
Saul~\cite{KeplerMapper2019}, along with the following specifications:
\begin{enumerate}
\item The projection function is the identity map, namely $f\colon X\to \mathbb{R}^4$
  simply sends a vector to itself. That is, since our data is not high-dimensional
  nor are we after any particular statistical features of the data cloud, there is no
  need to project it.

\item To cover the data (equivalently, the image of $f$), we use the standard
  Euclidean metric in $\R^4$
  and the default KeplerMapper cover procedure with Euclidean 4-dimensional cubes and
  the parameter $n=10$. This means that the projections of the data cloud onto each
  of the axes in $\R^4$ is covered by 10 overlapping intervals, and then each cube is
  formed as the cartesian product of those intervals. The degree of the overlap will
  be $\delta=8\%$. These values are chosen because empirically they appear to give
  the most informative Mapper graphs. These parameters will be modified slightly for
  some of the pictures in Section~\ref{sec:evolution}.

\item For the clustering algorithm, we use the default DBSCAN\footnote{See
    \href{https://scikit-learn.org/stable/modules/generated/sklearn.cluster.dbscan.html}{here}
    for more on DBSCAN.} clustering offered by KeplerMapper. The advantage of DBSCAN
  is that it identifies clusters of any shape (unlike other methods which only look
  for convex clusters).
\end{enumerate}

The colors in our Mapper graphs do not have any mathematical meaning. The nodes are
colored according to how the data is ordered, and this is done based on the
geographic information, so that nearby counties are colored similarly. We can thus
use the colors to add coherency among graphs and to help distinguish geographic
locations.



\begin{figure}[h]
  \includegraphics[width=0.8\linewidth]{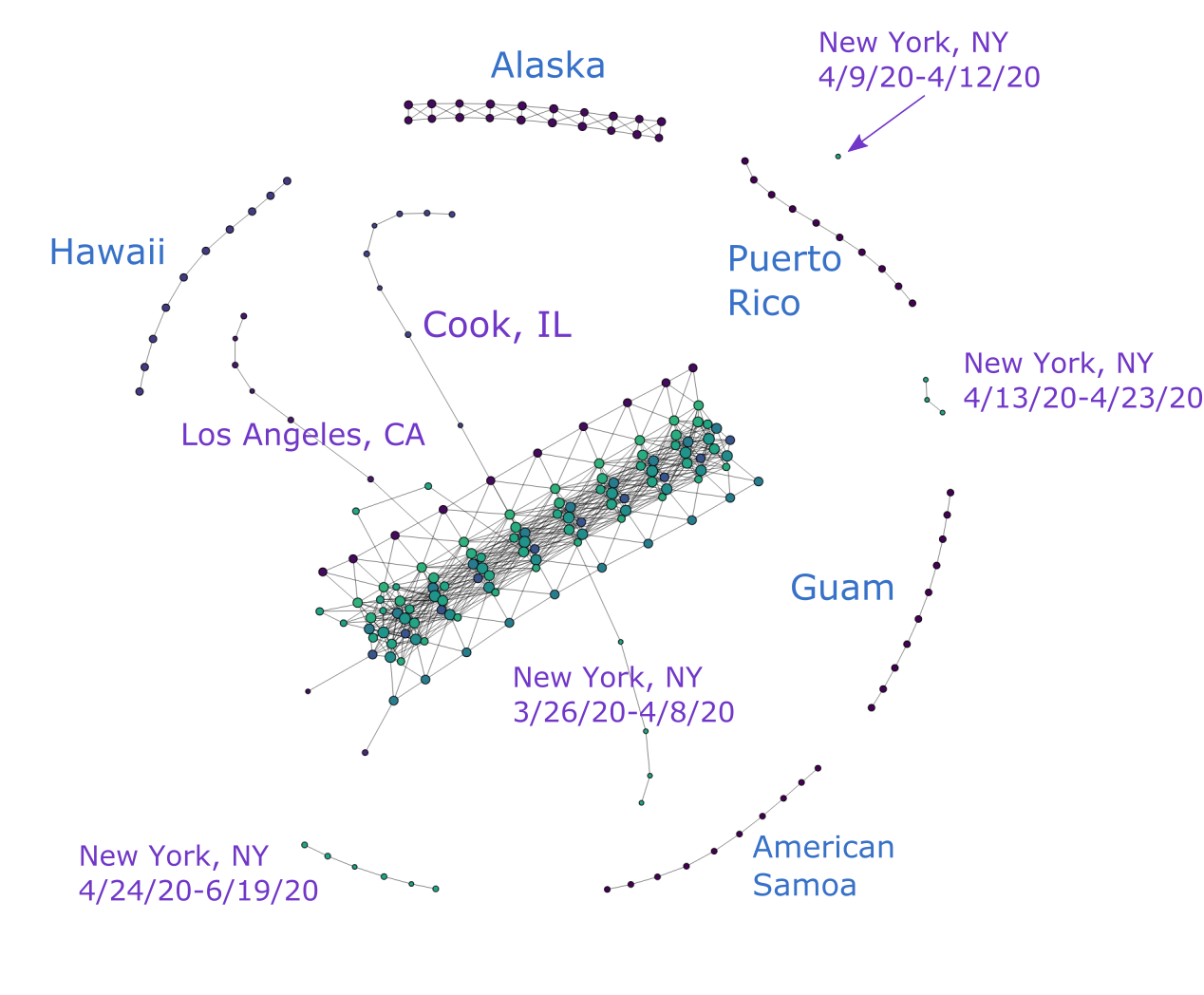}
  \caption{COVID-19 cases reported in the U.S. as of 6/19/20.
    ($n = 10, \delta = 8\%$)}\label{fig:all-mapper}
\end{figure}

A representative Mapper visualization of our data can be found in
Figure~\ref{fig:all-mapper}. This is a visualization of the COVID-19 data from all
3155 U.S.~counties and territories as of 6/19/20.

\begin{figure}[h]
  \includegraphics[width=0.6\linewidth]{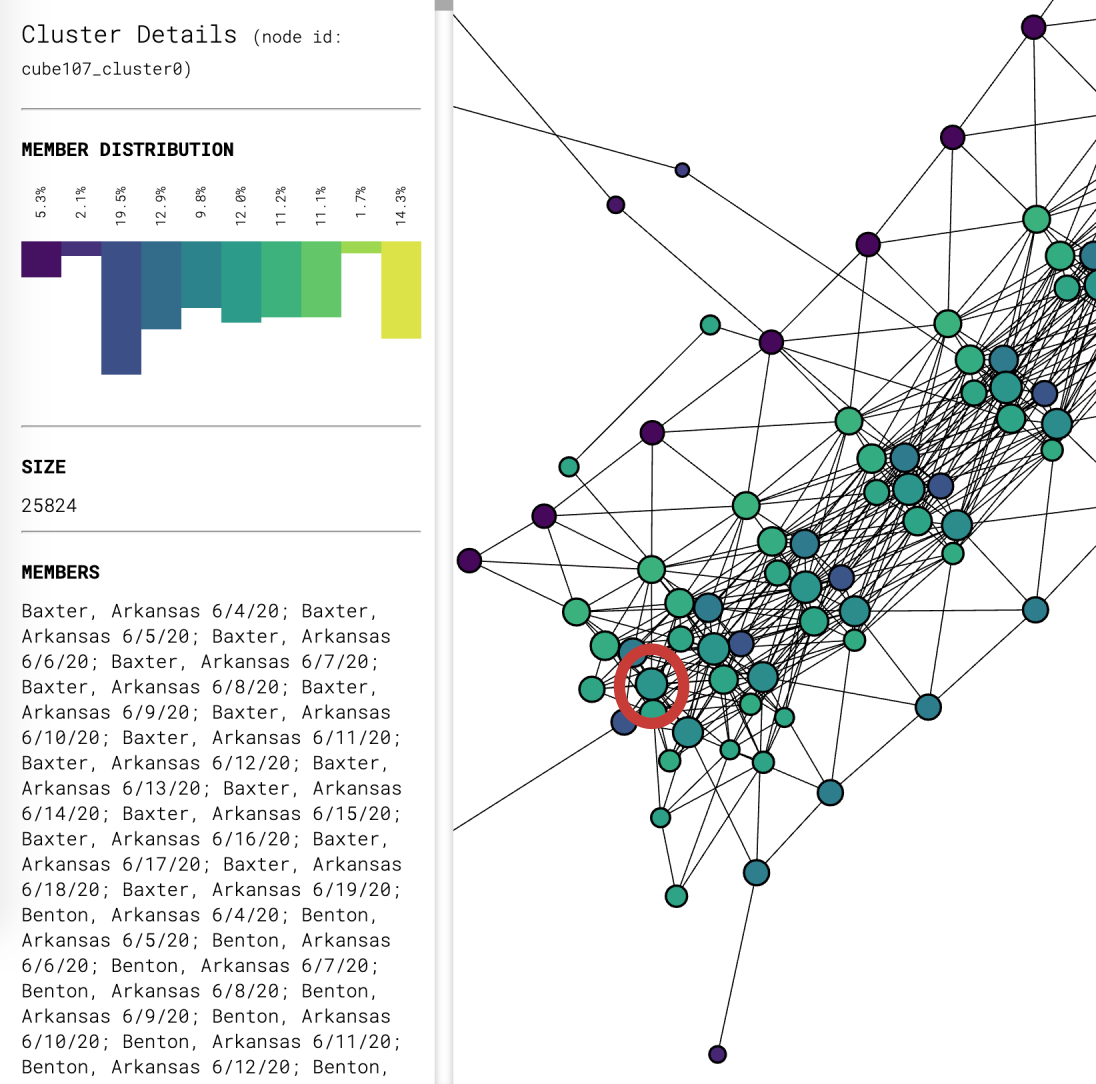}
  \caption{Breakdown of the data inside the node circled in red. Only the beginning
  of the list of the 25,824 data points in this node is shown.}\label{fig:node1}
\end{figure}

\begin{figure}[h]
  \includegraphics[width=0.6\linewidth]{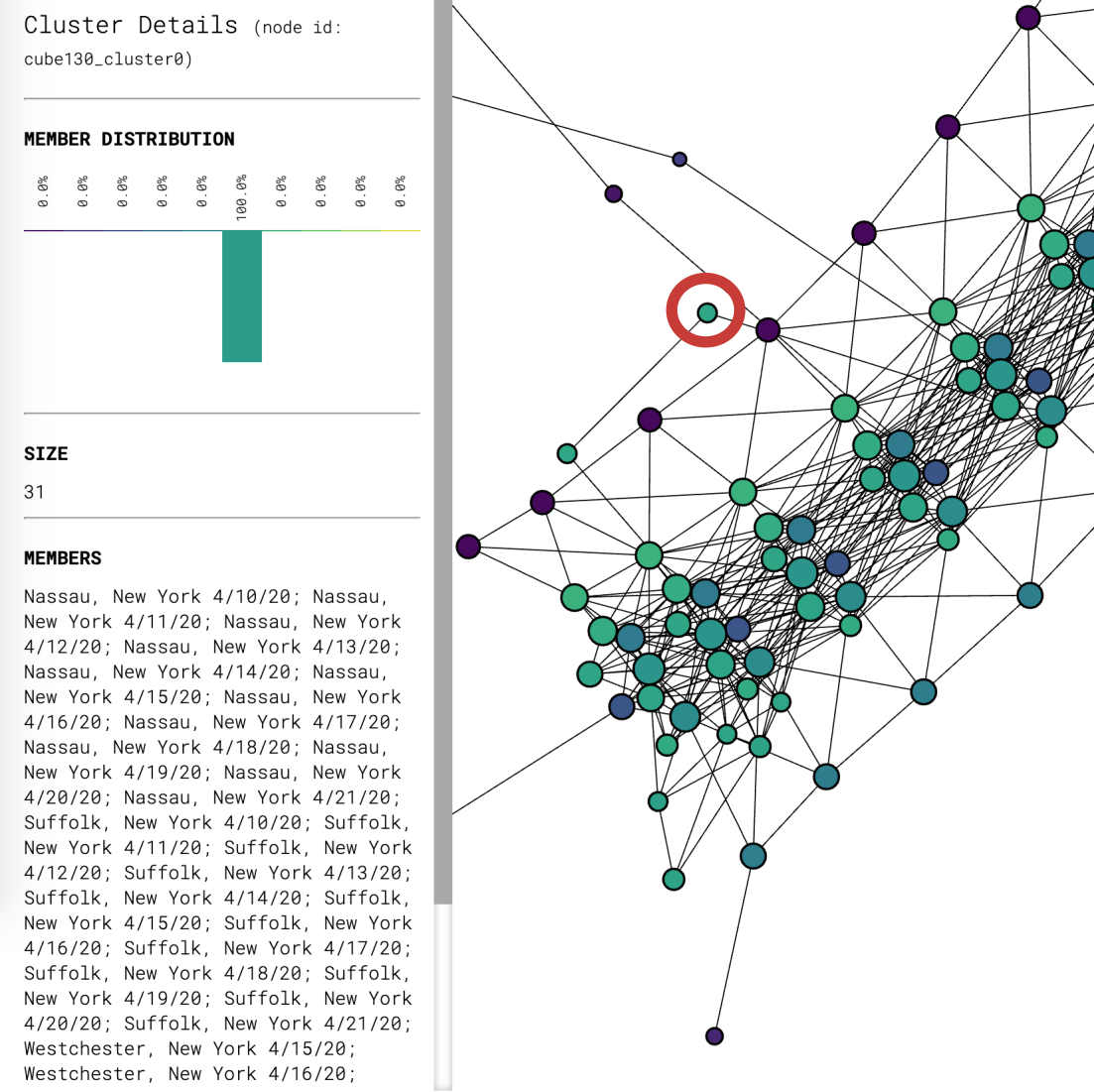}
  \caption{Breakdown of the data inside the node circled in red. The node contains 31
    data points, all of which come from New York State.}\label{fig:node2}
\end{figure}

Figures~\ref{fig:node1} and~\ref{fig:node2} show partial content of two
representative nodes. The first node contains 25,824 data points. The geographic
diversity of the counties in this node is evident from the number of bars of
different colors in the top left. In contrast, the second node only has 31 data
points, all of which come from New York State. Additionally, the size of each node is
also indicative of the number of data points it contains.

\subsubsection{Filtering}\label{ssec:filter}%


As mentioned in Section~\ref{sssec:normalization}, each of the four vectors
consisting of a particular coordinate of each data point is scaled to unit norm. The
resulting normalization for a data point is determined by the relative numerical
significance of each coordinate within their own coordinate peers. This normalization
is hence substantially affected by outliers in the data set, which can obscure a
significant amount of data that would otherwise be numerically distinguishable. For
example, looking at Figure~\ref{fig:all-mapper}, all counties in New York State, Los
Angeles County in California and Cook County in Illinois have disproportionately
large numbers of cumulative COVID-19 cases. They significantly increase the range of
the data and are thus given more weight on the normalization process. As a result,
their presence in the point cloud reduces the numerical significance of other data
points and makes them rather indistinguishable in the Mapper graph.

\begin{figure}[h]
  \includegraphics[width=0.9\linewidth]{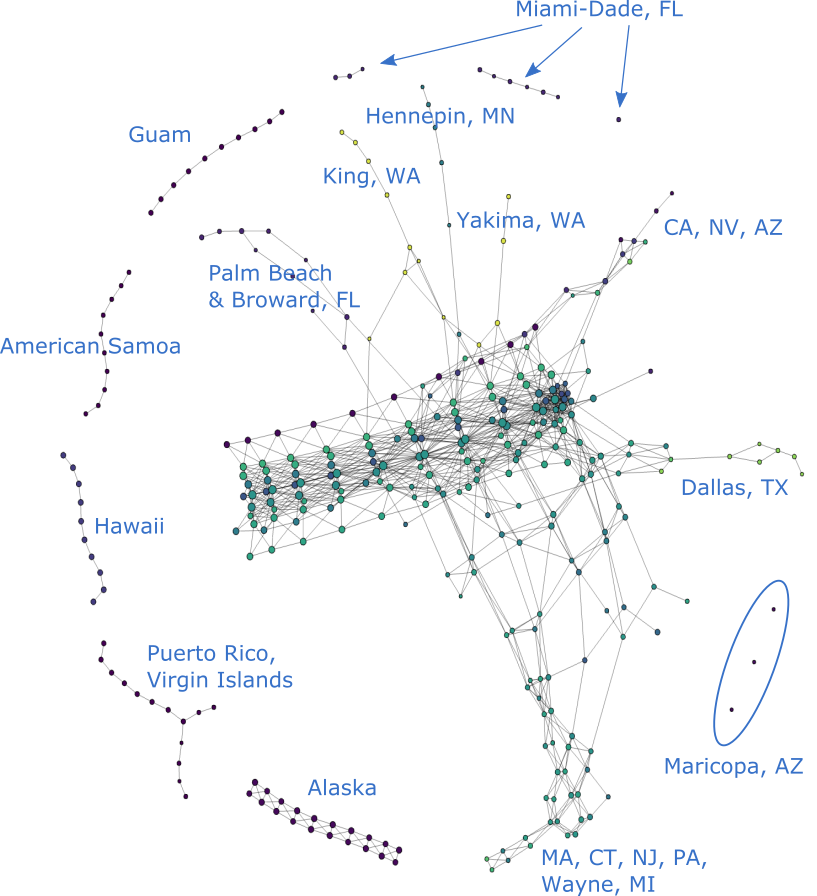}
  \caption{COVID-19 cases reported in the U.S. as of 6/19/20, excluding data from the
    entire New York State, Los Angeles County, CA and Cook County, IL.\@
    ($n = 10, \delta = 8\%$)}~\label{fig:filtered-mapper}%
\end{figure}

To solve this issue, we filter our data set to produce different graphs either with
or without these outliers. In the rest of the paper, we refer to data sets that
include the aforementioned outlier counties as ``unfiltered'' and the ones without as
``filtered''. For instance, Figure~\ref{fig:all-mapper} makes it visually obvious
that New York, Cook and Los Angeles ``stand out'' in the form of branches in
comparison with rest of the data that congregate in the main trunk. On the other
hand, Figure~\ref{fig:filtered-mapper} provides the Mapper graph representing the
same data set, but filtered to exclude data from Los Angeles, Cook, IL and the
entire New York state. It is evident from this graph that more branches emerge as a
result of filtering. This provides more helpful information on places that are not
top-ranked in terms of total COVID-19 cases but still display worrisome trends or
regional significance, such as King, WA and Maricopa, AZ.

The rest of the paper works with both unfiltered and, for the most part, filtered
data sets. We mostly use the former for New York-related analysis and the latter for
other prominent U.S.~COVID-19 clusters like the Maricopa County in Arizona.

For the filtered and unfiltered graphs that contain more recent dates (up to
7/24/20), see Figures~\ref{fig:update-all} and~\ref{fig:update-filtered}.


\section{Results}\label{sec:results}


In this section, we introduce several significant topological features that could be
observed from the Mapper graph, including components and branches, and illustrate
what could be inferred from the presence and structure of these features.

\subsection{Components}\label{ssec:components}
It is clear from Figure~\ref{fig:all-mapper} that the original point cloud clusters
into several disjoint components under the Mapper algorithm. Most of the data lives
in the main ``trunk,'' but there are also several smaller isolated components. These
disjoint components are formed either by a single isolated node or by a group of
nodes connected by edges. Recall
that an edge is created between two nodes when there is a data point living in both,
and that a node is formed by covering the projections of the point cloud to each axis
in $\R^4$ with overlapping, equal-length intervals and then clustering the data
points within each 4-dimensional cube determined by these intervals. Connected
clusters of nodes are thus formed when the data in each node displays close
similarity or proximity, and disjoint components emerge when they are numerically
dissimilar, so that the data points are ``far away,'' i.e.~there are gaps between the
projections of data points into any of the four dimensions, resulting in empty
overlaps between covers. Significant variations in any of the four coordinates among
data points therefore break the nodes or clusters of nodes into disjoint components.

Because the projection of the point cloud is linear along the time dimension before
the normalization, this distribution does not result in empty overlaps when covers
are applied. Therefore, instead of breaking the Mapper into isolated components, the
time dimension rather accounts in part for the connectedness features of the Mapper
and, furthermore, to the internal distribution of nodes within each
component. Therefore, we can usually explain the existence of isolated components by
variations in the other three coordinates, which concern two factors:
\begin{itemize}
\item[(a)] Geographic variations of the counties represented by the data
points; and
\item[(b)] Variations in the number of cumulative confirmed cases reported by each
county.
\end{itemize}
We will discuss the ramifications of these two factors below in
Sections~\ref{ssec:geography} and~\ref{ssec:cases}. Before that, we will say more
about how the time component plays a role in keeping the nodes connected and in
shaping the internal structure of most components.


\subsubsection{Time}\label{ssec:time}%


The time coordinate is unique in the sense that its values
follow a linear progression before normalization. Because of this linearity,
normalized coordinates stay relatively dense and empty overlaps will thus almost
never emerge when covers are applied to the projection of the data into the time
axis, notwithstanding most choices of $n$ and $\delta$. This implies that time hardly
explains the disconnectedness of the graph but in fact produces a considerable number
of the edges between nodes and clusters.

\begin{figure}[h]
  \includegraphics[width=0.6\linewidth]{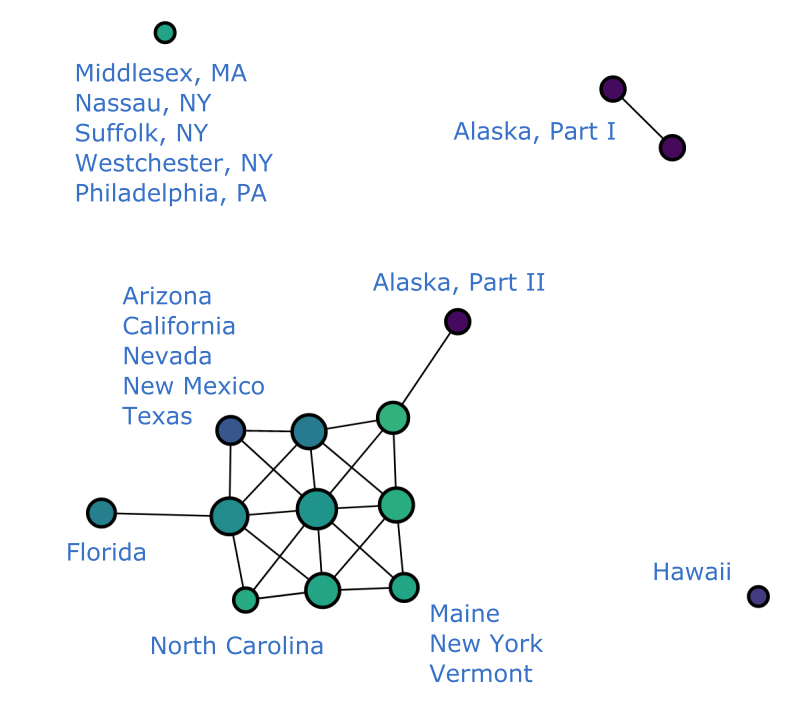}
  \caption{Cumulative COVID-19 cases reported in the U.S. on 7/1/20. ($n = 10, \delta
    = 8\%$)}
  \caption*{\textit{Note:} Alaska, Part I houses the following Alaskan
    counties: Anchorage, Bethel, Denali, Fairbanks North Star, Kusilvak,
    Matanuska-Susitna, Nome, North Slope, Northwest Arctic, Southeast Fairbanks,
    Valdez-Cordova, Yukon-Koyukuk, Aleutians East, Bristol Bay, Dillingham, Kenai
    Peninsula, Kodiak Island, Lake and Peninsul, Yakutat, while Alaska, while Part II
    houses Aleutians West, Haines, Hoonah-Angoon, Juneau, Ketchikan Gateway,
    Petersburg, Prince of Wales-Hyder, Sitka, Skagway, and Wrangell.}\label{fig:single}
\end{figure}

\begin{figure}[h]
  \includegraphics[width=0.8\linewidth]{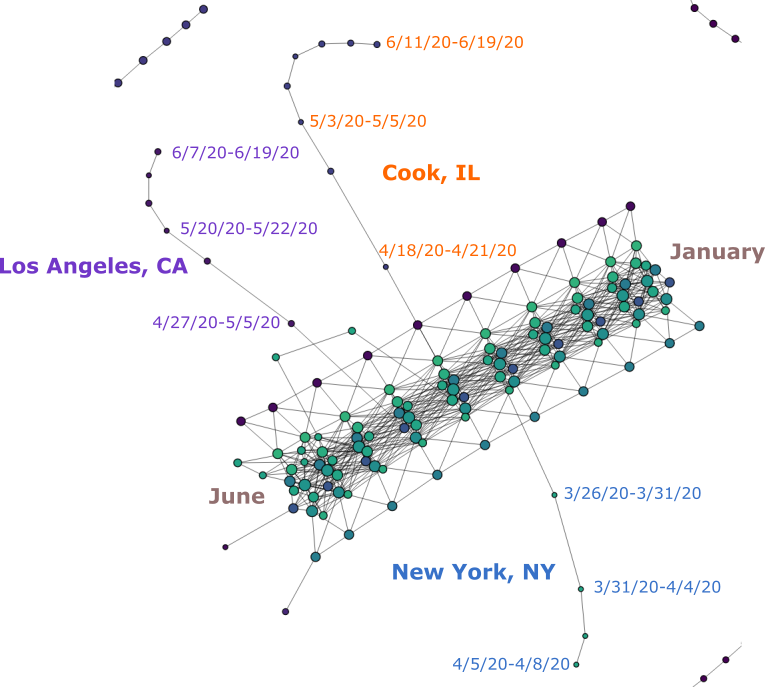}
  \caption{Main trunk and branches of Figure~\ref{fig:all-mapper} showing time
    progression.}\label{fig:all-dates}
\end{figure}

Moreover, when covered by equal-length intervals, this dense distribution results in
highly predictable, segmented internal structures within most clusters in the Mapper
graph, so that the main trunk and most other substructures could even be seen as
having been formed by weaving together clusters or nodes with data from successive
time segments. To see this, let us first remove the time coordinate and produce a
Mapper graph with data from a single day as in Figure~\ref{fig:single}. Without the
time component, we can clearly observe the impact of geography on the shape of the
graph. If we compare this graph with Figure~\ref{fig:all-dates}, which is generated
with the time coordinate, we see that the latter graph could be regarded as
having been formed by horizontally juxtaposing and connecting graphs produced on
consecutive dates, each looking like Figure~\ref{fig:single}, and then applying the
clustering algorithm along the time axis to further cluster the points.

The result, as we can observe in Figure~\ref{fig:all-dates}, is that our Mapper graph
displays the evolution of cases through the progression of time in an intuitive
manner, where the trunk and most other substructures appear to have ``grown'' in
accordance with time data like the trunk of a tree, building up on data from the
first few days in our data set. The right side of the trunk contains most of the
U.S. counties at the beginning of the time period in January, and the time progresses
toward the left. The left end of the trunk contains the data for June. The
``branches'' that stick out of the main trunk also follow this pattern. The nodes
that make up the branches in Figure~\ref{fig:all-dates} are clearly arranged in
accordance with the progression of time. Hence each branch represents the development
of COVID-19 cases within a certain period of time. We will study these branches in
depth in Section~\ref{ssec:branches}.

\subsubsection{Geography}\label{ssec:geography}%


Geographic information is an important factor that affects the distribution of the
data throughout the Mapper graph and is readily reflected through this visualization.
For example, the nodes representing Hawaii, Alaska, Guam, Puerto Rico and American
Samoa in Figure~\ref{fig:all-mapper} naturally form independent clusters. This is
explained by their geographic separation from the U.S.~mainland, which is for the
most part represented in the Mapper graph as the main trunk. Specifically, the
geographic separation for these regions is reflected in the data in the relative
numerical disparity in the first two coordinates. With appropriate values of $n$ and
$\delta$, this numerical disparity will give rise to disjoint clusters lying outside
of the main trunk in the Mapper graph, as they do in Figure~\ref{fig:all-mapper}.

The geography is also more transparent when we remove the time coordinate and produce
a Mapper graph for a single day only, as we do in Figure~\ref{fig:single}. The
grouping of data points into clusters highly correlates with U.S.~geography. However,
we can also spot an isolated node in the graph representing several counties in New
York, Pennsylvania and Massachusetts. This result should clearly not be attributed to
geography and we thus know that geographic information is not deterministic to the
shape of the Mapper graph. Since we have already removed the time component from our
data before generating this graph, the only coordinate left that could explain this
distribution is the number of cases reported.


\subsubsection{Number of COVID-19 cases}\label{ssec:cases}%


As explained in Section~\ref{ssec:data}, the third coordinate of our data points
records the cumulative number of COVID-19 cases reported daily from each county. In
comparison with geographic and temporal information, this coordinate usually has the
largest standard deviation among the four features.\footnote{For example, for a point
cloud with data up to 6/21/20, the standard deviations for the four coordinates are
calculated to be 0.0002336, 0.00022721, 0.0014227 and 0.00072409, respectively, after
normalization.} Before normalization, its numerical value can range from a few
hundred for less-affected areas to several hundred thousand for hot-spots like the
New York City, or, more recently, some areas in Florida and Texas. Therefore, this
coordinate also adds the most variability into the shape of the Mapper graph.

This is mostly evident in two aspects. Similar to what happens with the geographic
information, a large disparity in the numerical values of this coordinate can
distribute data points into disjoint components outside of the main trunk. For
example, we see in Figure~\ref{fig:single} that there is an isolated node in the
upper left corner whose presence could not be explained by geographic isolation. In
fact, this node houses several counties with the highest numbers of cumulative
COVID-19 cases reported by the time this graph was generated. The Mapper
representations of such counties thus correspondingly stand out from those of the
places that report only a mediocre number of cases. This type of distribution is also
observed in ordinary Mapper graphs like Figure~\ref{fig:all-mapper} where temporal
data is included.

For the most part, however, the elevations of COVID-19 cases are reflected through
the emergence of ``branches'' sticking out of the main trunk. Under this point of
view, the free-floating clusters or nodes discussed above can be seen as fragmented
segments of an otherwise connected branch. We will examine these branches in more
detail next.


\subsection{Branches}\label{ssec:branches}%


We see in Figure~\ref{fig:all-mapper} that the data for Los Angeles, CA, Cook, IL and
New York, NY presents itself in the form of chains of nodes branching off the main
trunk. These three counties ranked top three in terms of COVID-19 cases reported by
the time this graph was generated. As discussed in previous sections, because
geographic and temporal data have relatively small standard deviation, they cause the
graph to stay connected in these dimensions rather than producing features like these
branches. This is evident in the ``branch-free'' part of the trunk, which represents
data from early days of the outbreak when there is no significant disparity in the
numbers of cases reported from different regions. It is therefore the high number of
cases reported in these places that forces the nodes representing these counties to
depart from the ones representing their geographic neighbors.

With appropriate choices of $n$ and $\delta$, we can thus produce Mapper graphs where
regions with relatively large numbers of COVID-19 cases are no longer bound to others
through edges in the geographic dimensions but remain connected solely in the time
dimension, resulting in the branching feature that we see in
Figure~\ref{fig:all-mapper}.\footnote{Branches like these are sometimes called
``flares'' in the literature~\cite{EHIO:MapperFirms}. Our choice of terminology comes
from the fact that we also have an evident ``trunk.'' In addition, ``flaring''
already has a meaning in the context of a spread of a virus.} Hence, the growth of
these branches tracks the incremental increase of cases in these counties.

Additionally, because different segments of the main trunk are built with their own
timestamps as discussed in Section~\ref{ssec:time}, the place where each branch
starts to grow is therefore indicative of the onset of the outbreak that it
represents. That is to say, a branch that emerges later in the trunk signifies a more
recent outbreak. In this way, we can see that the Mapper graphs encodes the
development of the pandemic in these hot-spots in an intuitive manner. In the
following sections, we will offer several case studies on various branches of
different shape and elaborate on what could be learned from them.

\subsubsection{Segmented branches: New York and others}
As we just explained, the emergence of branches signifies potential or existing
COVID-19 hot-spots. However, some hot-spots like New York are more prominent than
others; this empirical phenomenon also has a representation in the Mapper graph.
Namely, looking at Figure~\ref{fig:all-mapper}, what we observe is that the branch
representing data from the New York County is in fact broken into several segments,
with each segment representing COVID-19 cases reported during a certain period of
time. This distribution can be regarded as reflecting several stages in the
development of the pandemic in New York during which several dramatic spikes of cases
occurred.

\begin{figure}[h]
  \includegraphics[width=\linewidth]{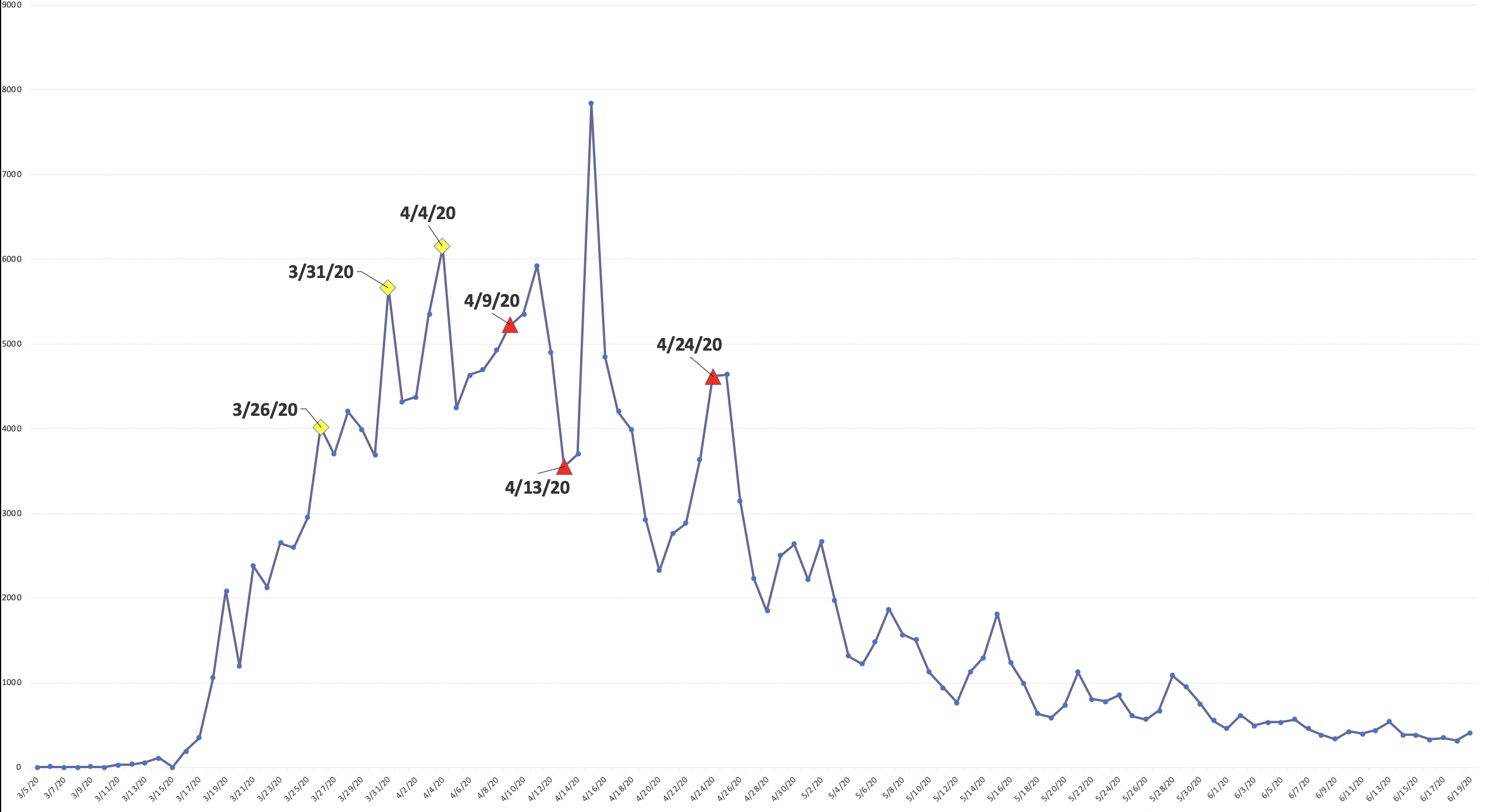}
  \caption{Daily new cases reported in New York, NY from 3/5/20 to
    6/19/20.}\label{fig:daily_new_ny}
\end{figure}

In particular, 4/9/20, 4/13/20, and 4/24/20 are the critical dates underlying these
disruptions in the branch. As is visible in Figure~\ref{fig:daily_new_ny}, these
dates (labeled in red) correspond to either spikes in the number of daily new cases
reported in New York County or a nadir before the next peak. Hence, the disruptions
in the branch usually occur as a result of an extended period of elevated daily
incidence,
which forces the projections of the succeeding series of data points along the
``number of cases'' dimension to land in a hypercube far enough so that it breaks the
continuity established by densely populated data points along the time dimension.
Additionally, we notice that earlier peaks of daily new cases did not break up the
branch. Instead, they contribute to the growth in the ``length'' of the branch by
adding a new node. That is, several such peaks that occurred on 3/26/20, 3/31/20, and
4/4/20 (labeled in yellow) correspond to the critical dates that cause a new node to
be added to the branch.
Therefore, many noteworthy dates in the development of the pandemic in New York
County are reflected in the graph either in the form of new nodes or disruptions of
the branch. Several snapshots of the dynamic process through which segments of this
branch gradually emerged can be found in Figure~\ref{fig:evo-ny}.

Similarly, in the Mapper graphs generated from filtered data sets, some prominent
COVID-19 hot-spots are also exposed in the form of segmented branches. They are
Miami-Dade, FL and Maricopa, AZ, and can be seen in Figure~\ref{fig:filtered-mapper}.
Similarly to New York, these places also experienced aggressive growths in daily
incidence. Broken branches therefore generally symbolize high levels of daily growths
and signify alarming trends in the regions they represent.


\subsubsection{Branch complex}~\label{ssec:complex}
A branch complex is formed when a number of closely-situated counties report similar
high numbers of cases and display elevated daily incidence. In this case, each county
is represented as a branch that departs from the main trunk. However, because of the
geographic and temporal proximity of these outbreaks, nodes in their respective
branches tend to be connected through the geographic dimension as well. This explains
the various entangled branching structures emanating from the main trunk in
Figure~\ref{fig:filtered-mapper}.\footnote{This structure is unlikely to emerge in
Mapper graphs for unfiltered data because New York data obscures the other minor
hot-spots and makes them indistinguishable (see Section~\ref{sssec:normalization}).}
For example, one can see the complex formed by a number of counties in the Northeast
and Wayne, Michigan as well as a less complicated one formed by counties in
California, Nevada, and Arizona. Another small complex is formed by two counties in
Florida, i.e.~Palm Beach and Broward.

Because of their prominence, these complexes tend to develop into trunks of their own
so that a resurgence of the pandemic in these existing hot-spots will be reflected in
the Mapper graph in the form of new branches that grow out of the branch complex
instead of the main trunk. On the other hand, the emergence of a new hot-spot in
nearby regions enlarges the complex by entangling a new branch into it.


\subsubsection{More recent hot-spots}


As mentioned in Section~\ref{ssec:branches}, the timing of the onsets of regional
flareups is indicated by the position of the resulting branches relative to the main
trunk. For example, we can see from Figure~\ref{fig:all-mapper} that the outbreak in
New York occurred prior to the ones in Cook and Los Angeles, because its branch
emerged earlier in time. We can similarly distinguish more recent clusters of
outbreaks from older ones in this way. For instance, it is clear in
Figure~\ref{fig:filtered-mapper} that the outbreaks in Dallas, Texas and Nevada
occurred after the ones in Palm Beach and Broward, Florida or in King, Washington as
their branches sit closer to the end of the trunk that corresponds to later dates.
They therefore represent a new generation of COVID-19 hot-spots, distinct from more
mature ones in Northeast U.S.~or Washington state.

\begin{figure}[h]
  \includegraphics[width=0.8\linewidth]{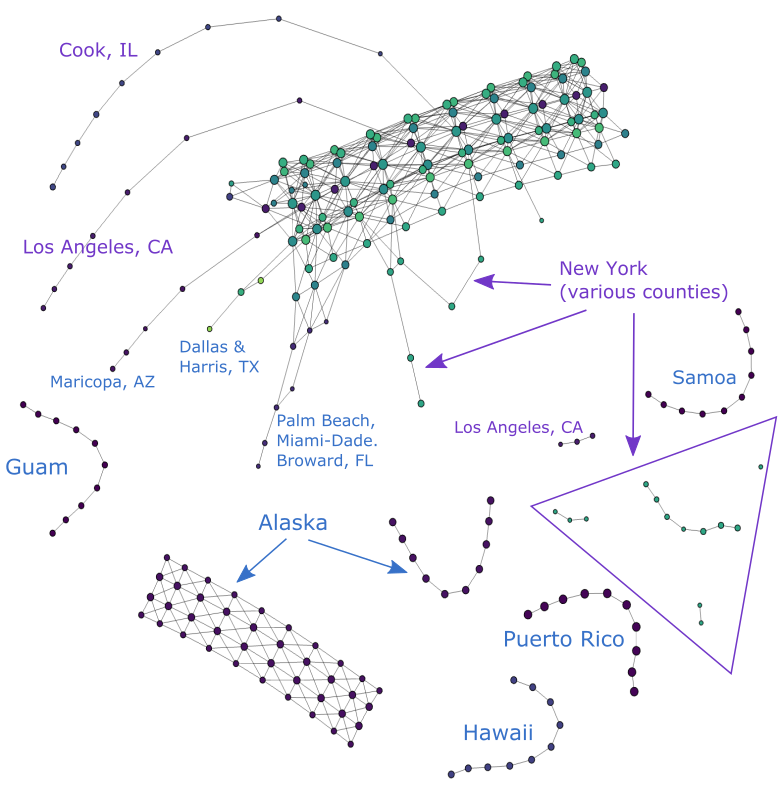}
  \caption{COVID-19 cases reported in the U.S. as of 7/24/20.
    ($n = 10, \delta = 8\%$)}\label{fig:update-all}
\end{figure}

\begin{figure}[h]
  \includegraphics[width=0.8\linewidth]{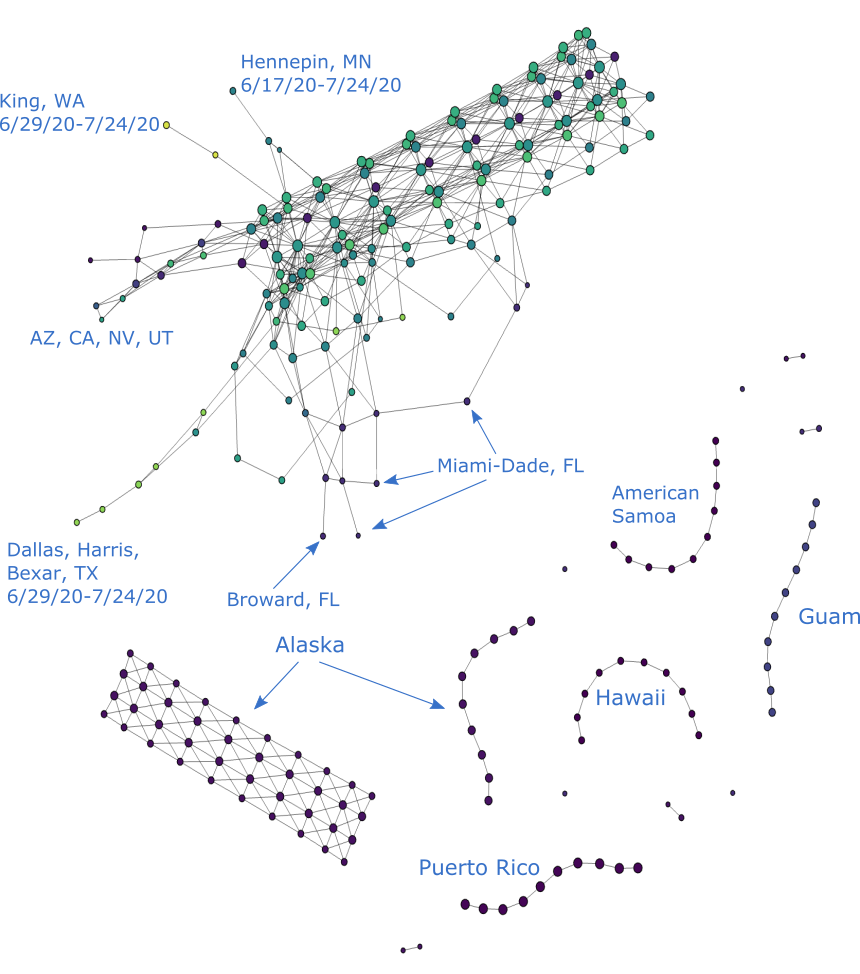}
  \caption{COVID-19 cases reported in the U.S. as of 7/24/20, excluding data from the
    entire New York State, Los Angeles, CA and Cook, IL.~The four isolated nodes are
    Miami-Dade, FL for 7/8-7/11, 7/12-7/15, 7/16-7/18, and 7/19-7/22. Three of the
    four components consisting of two nodes and one edge are Maricopa, AZ for
    7/1-7/5, 7/9-7/13, and 7/18-7/24. The fourth such component is Broward and
    Miami-Dade, FL for 7/5-7/21. ($n = 10, \delta = 8\%$)}\label{fig:update-filtered}
\end{figure}

To ensure the coherency in our analysis, we only studied graphs produced with data
collected prior to 6/20/20 in previous sections. For a more recent update in these
graphs, see Figures~\ref{fig:update-all} and~\ref{fig:update-filtered}. It is
noteworthy that these graphs make several new hot-spots readily visible. They include
Maricopa County in Arizona, Harris and Dallas Counties in Texas, and several places
in Alaska. Additionally, there appear to be resurgences in several existing hot-spots
so that their branches now become more distinguishable in the graph. Such places
include Miami-Dade and Broward Counties in Florida and several Californian counties.


From these updates, what we see is thus that, while embodying a coherent
developmental progress of the pandemic, Mapper graphs are also in themselves fluid
objects capable of change. We can get a more holistic picture of the entire progress
by generating Mapper graphs with real-time data and follow its path of evolution. In
the next section, we will discuss the evolution of Mapper graphs and the branches.


\subsection{Evolution of Mapper graphs}\label{sec:evolution}


Because our data incorporates temporal and geographic information, the resulting
Mapper graphs are capable of conveying useful information regarding the gradual
development of the COVID-19 pandemic across time and space. In
Section~\ref{ssec:branches}, we studied this by inspecting different clusters as well
as the branches sticking out of the main trunk of a particular Mapper graph. In this
section, we demonstrate how one can obtain a more holistic picture of the growth of
COVID-19 cases in the U.S.~by studying the evolution of these features across Mapper
graphs generated at different points in time.

\begin{figure}[h]
  \begin{subfigure}[b]{0.45\textwidth}
    \centering
    \includegraphics[width=\linewidth]{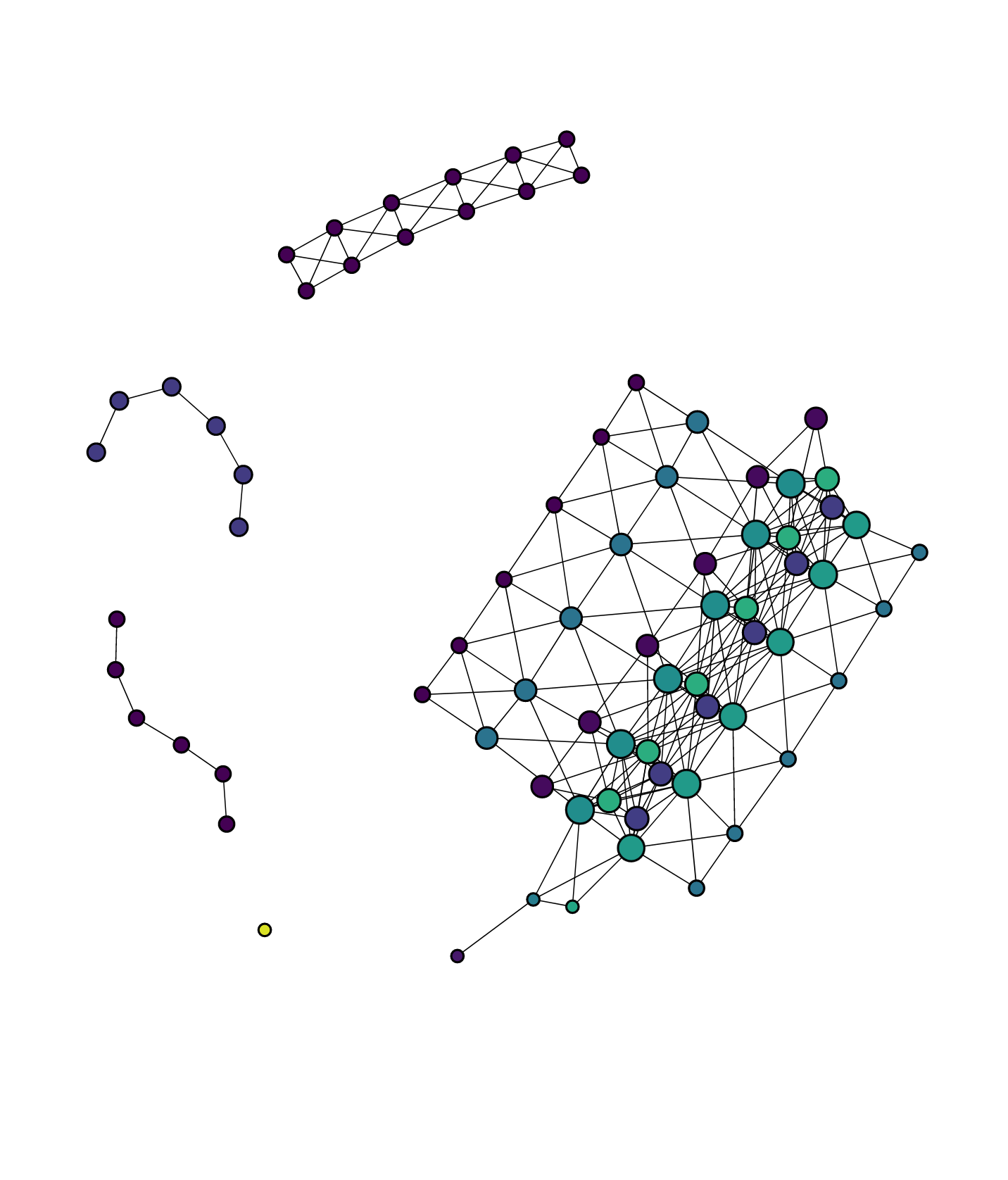}
    \caption{3/6/20 ($n = 6, \delta = 10\%$)}
  \end{subfigure}
  \begin{subfigure}[b]{0.45\textwidth}
    \centering
    \includegraphics[width=0.9\linewidth]{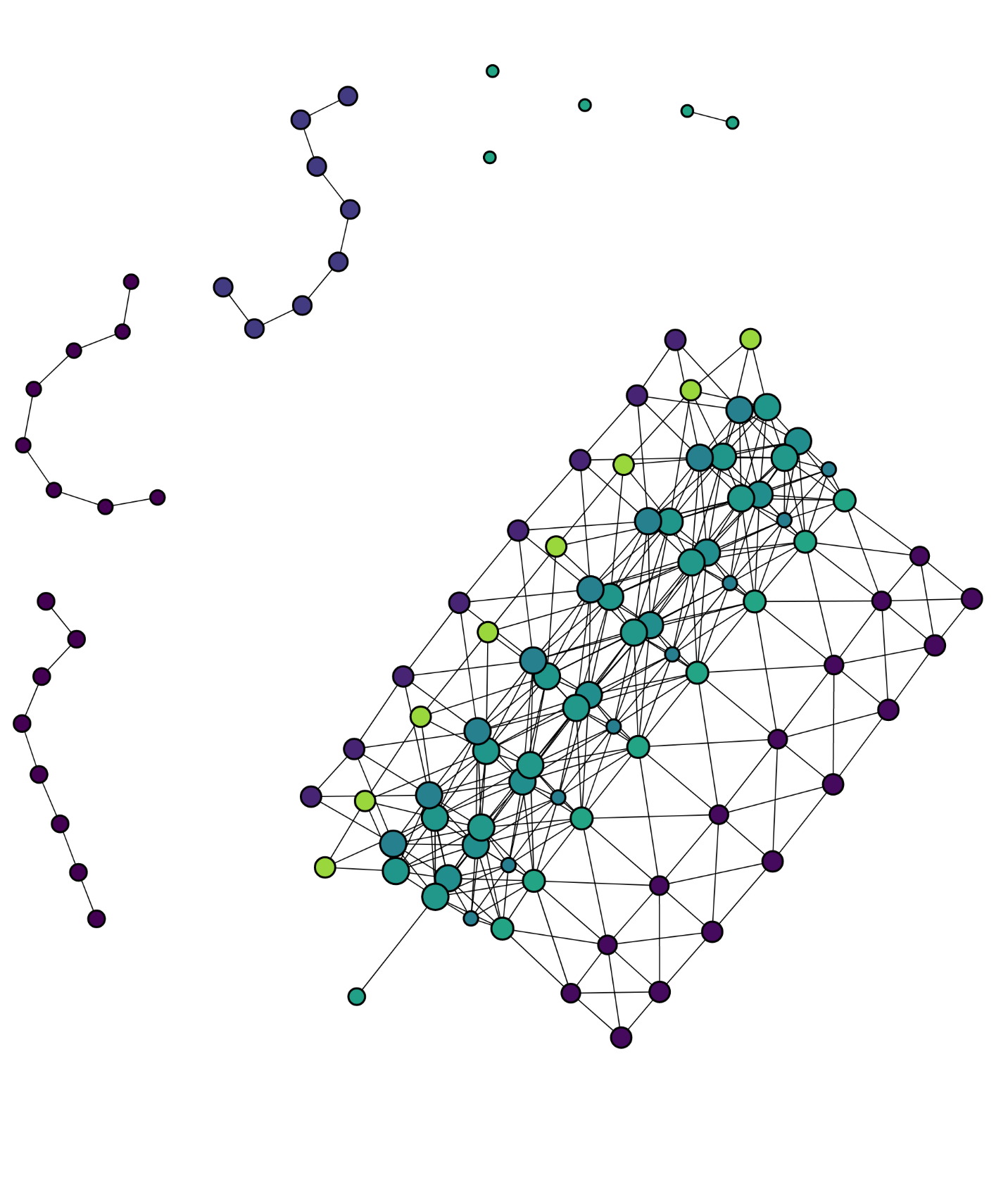}
    \caption{4/10/20 ($n = 8, \delta = 10\%$)}
  \end{subfigure}

  \begin{subfigure}[b]{0.45\textwidth}
    \centering
    \includegraphics[width=\linewidth]{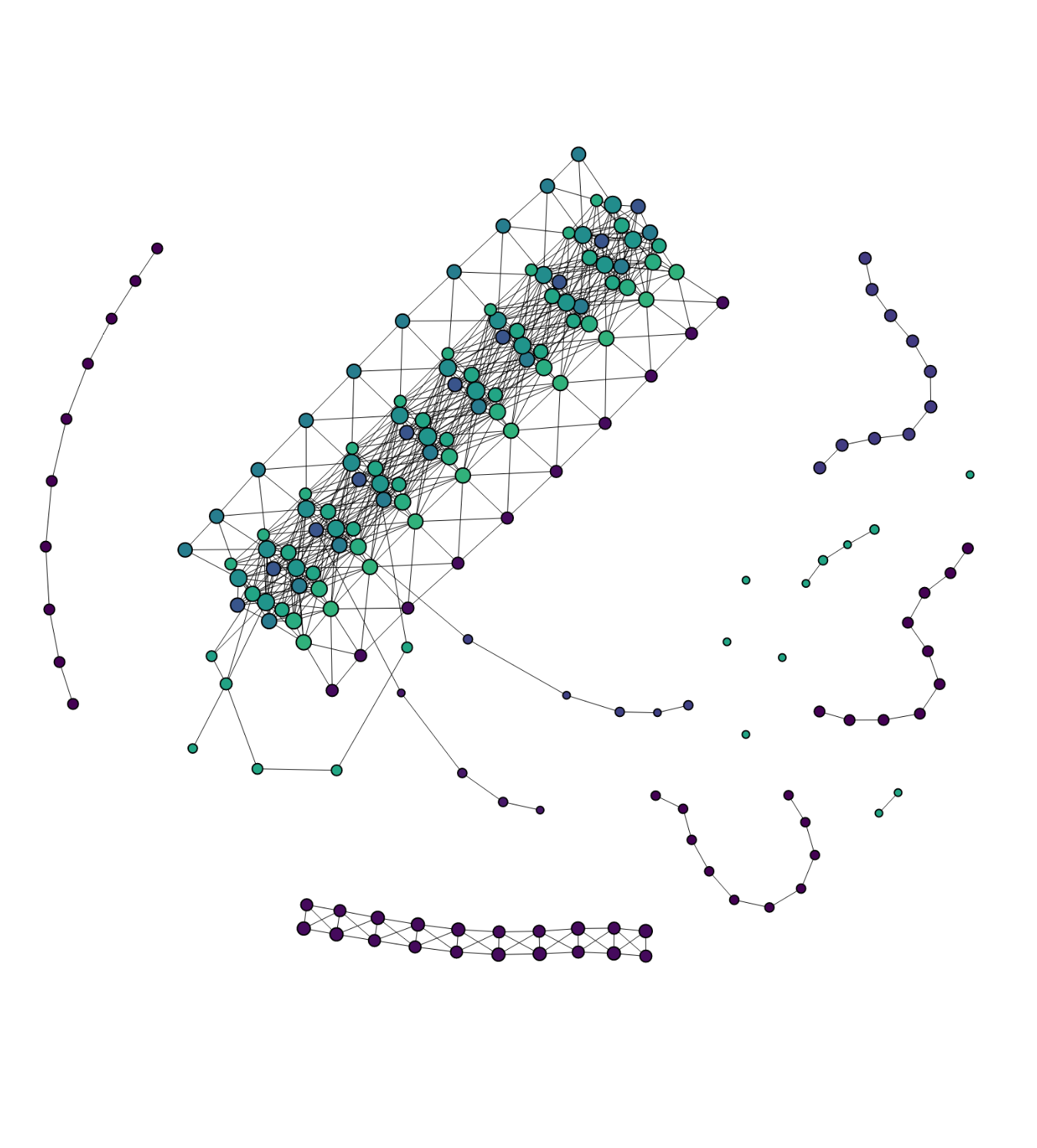}
    \caption{5/22/20 ($n = 10, \delta = 8\%$)}
  \end{subfigure}
  \begin{subfigure}[b]{0.45\textwidth}
    \centering
    \includegraphics[width=1.1\linewidth]{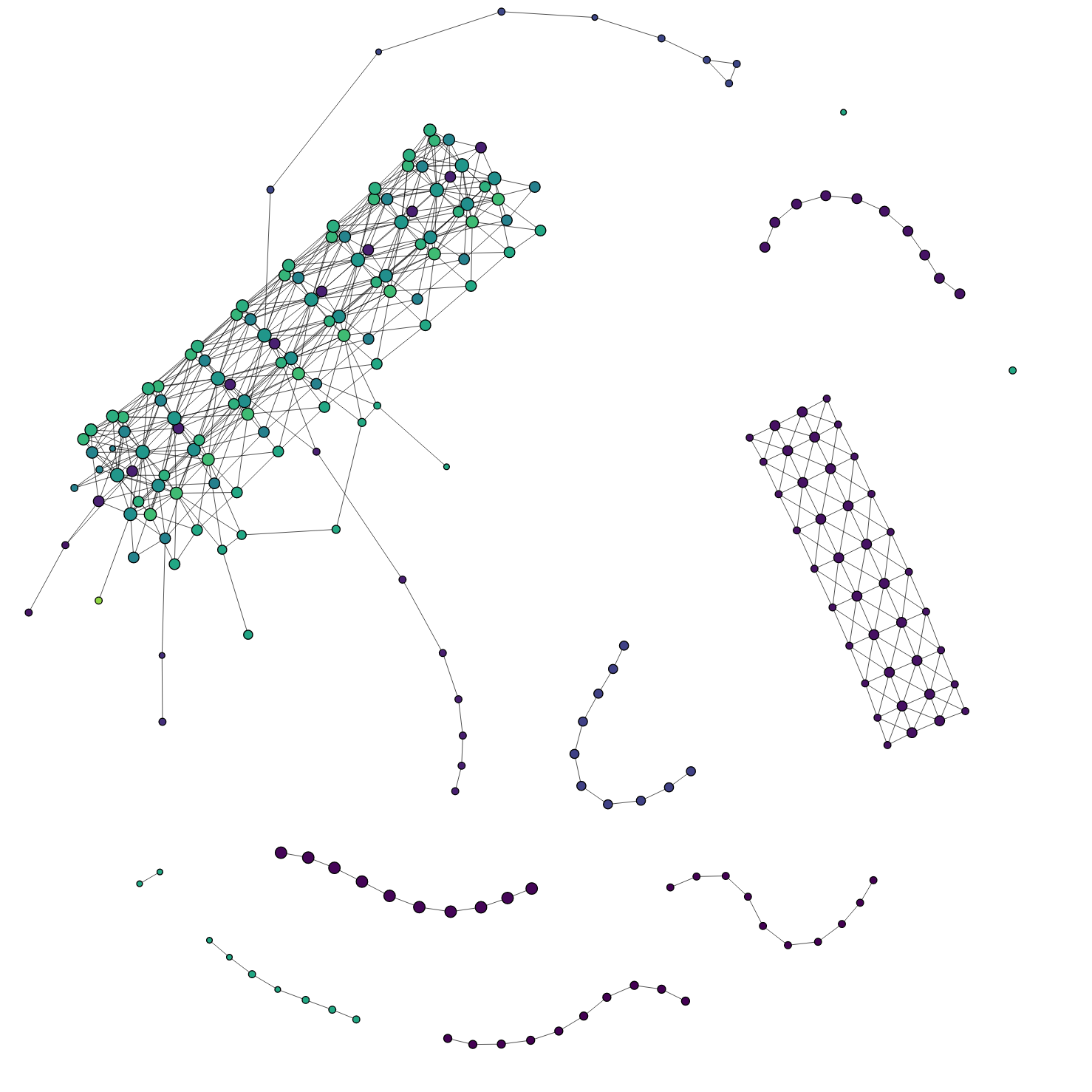}
    \caption{7/3/20 ($n = 10, \delta = 8\%$)}
  \end{subfigure}
  \caption{The evolution of Mapper graphs over time for unfiltered
  data.}~\label{fig:four-all}
  \end{figure}

\begin{figure}[h]
  \begin{subfigure}[b]{0.45\textwidth}
    \centering
    \includegraphics[width=\linewidth]{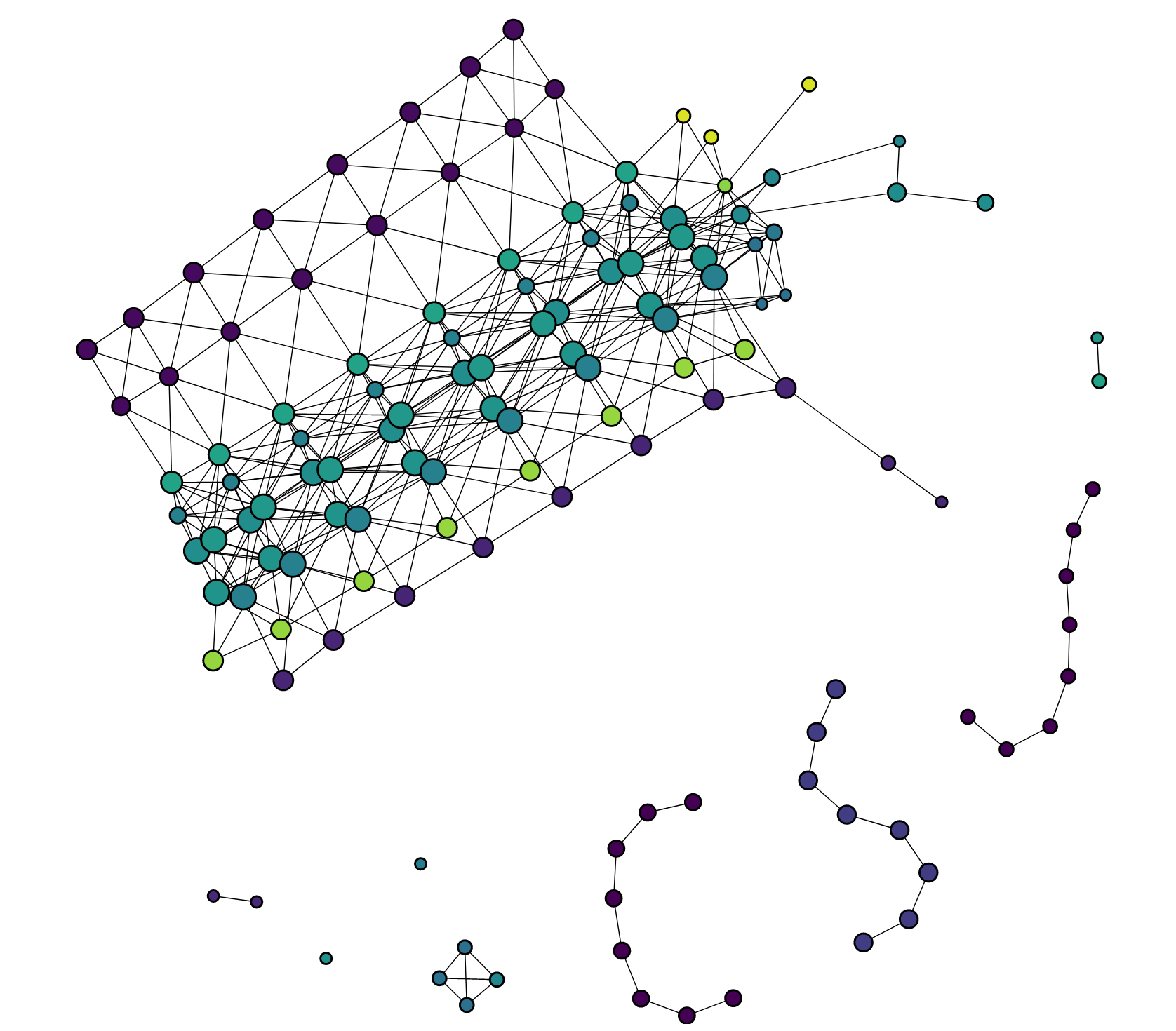}
    \caption{4/10/20 ($n = 8, \delta = 12\%$)}
  \end{subfigure}
  \begin{subfigure}[b]{0.45\textwidth}
    \centering
    \includegraphics[width=0.9\linewidth]{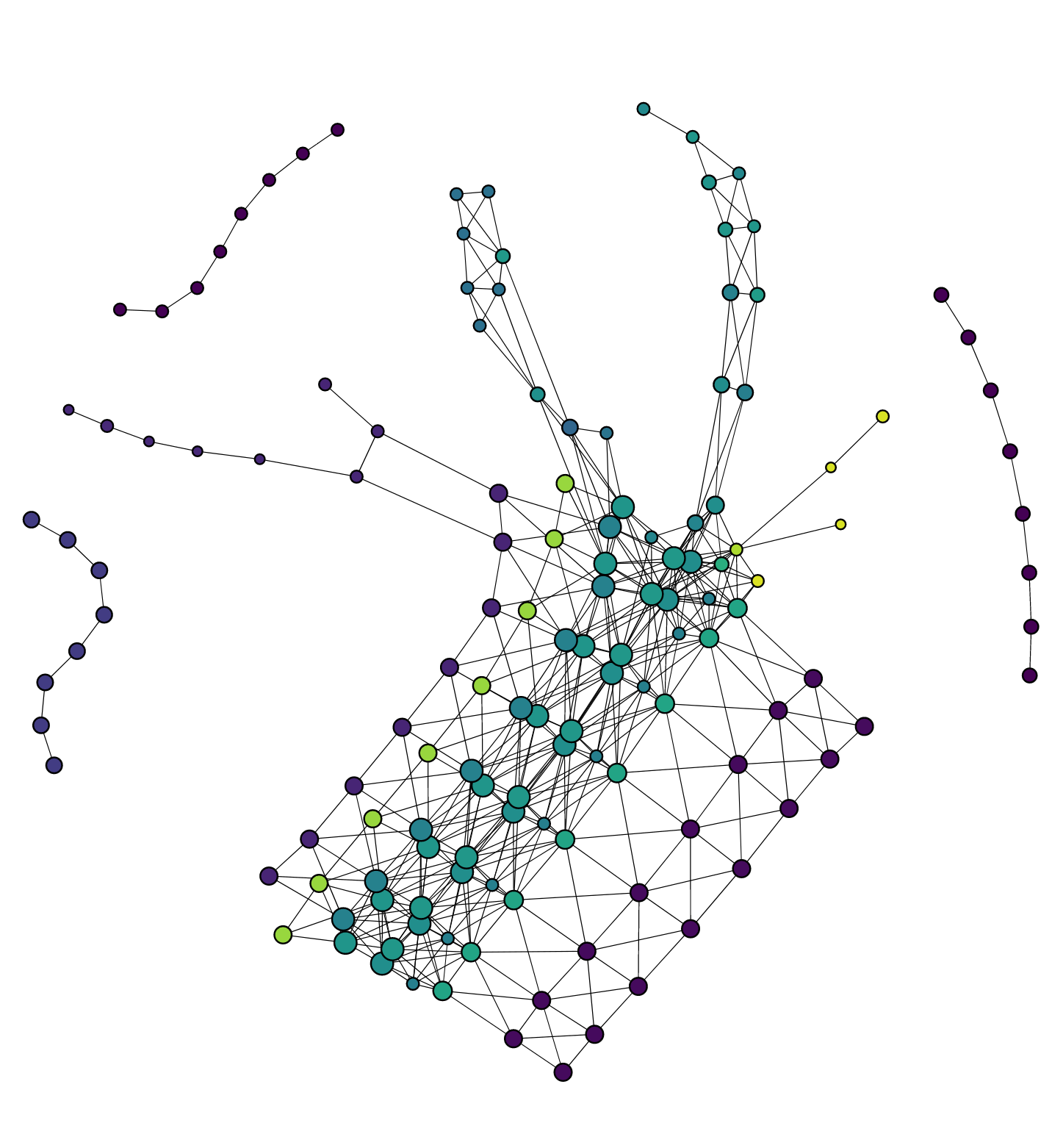}
    \caption{4/24/20 ($n = 8, \delta = 10\%$)}
  \end{subfigure}

  \begin{subfigure}[b]{0.45\textwidth}
    \centering
    \includegraphics[width=\linewidth]{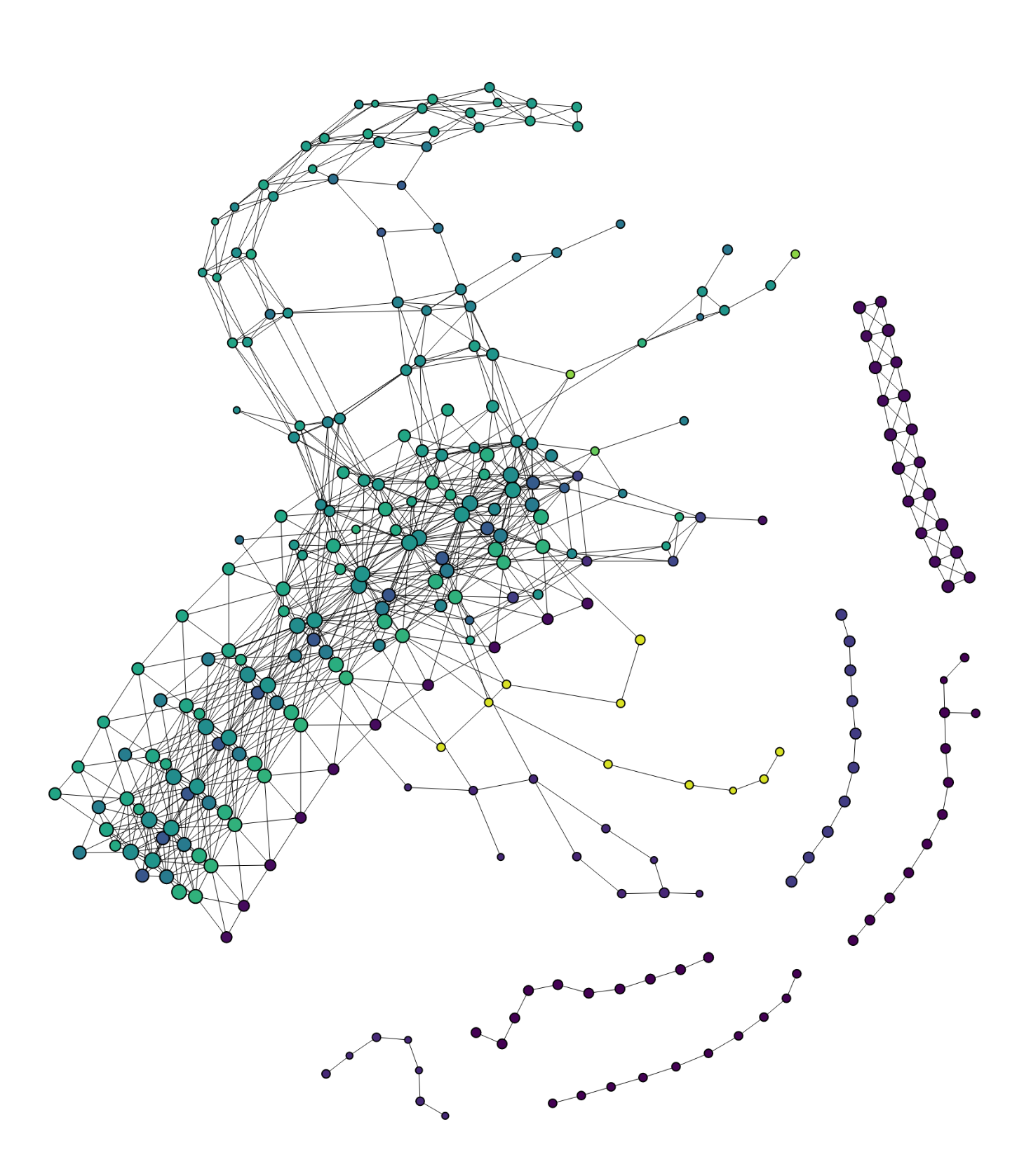}
    \caption{5/22/20 ($n = 10, \delta = 8\%$)}
  \end{subfigure}
  \begin{subfigure}[b]{0.45\textwidth}
    \centering
    \includegraphics[width=1.1\linewidth]{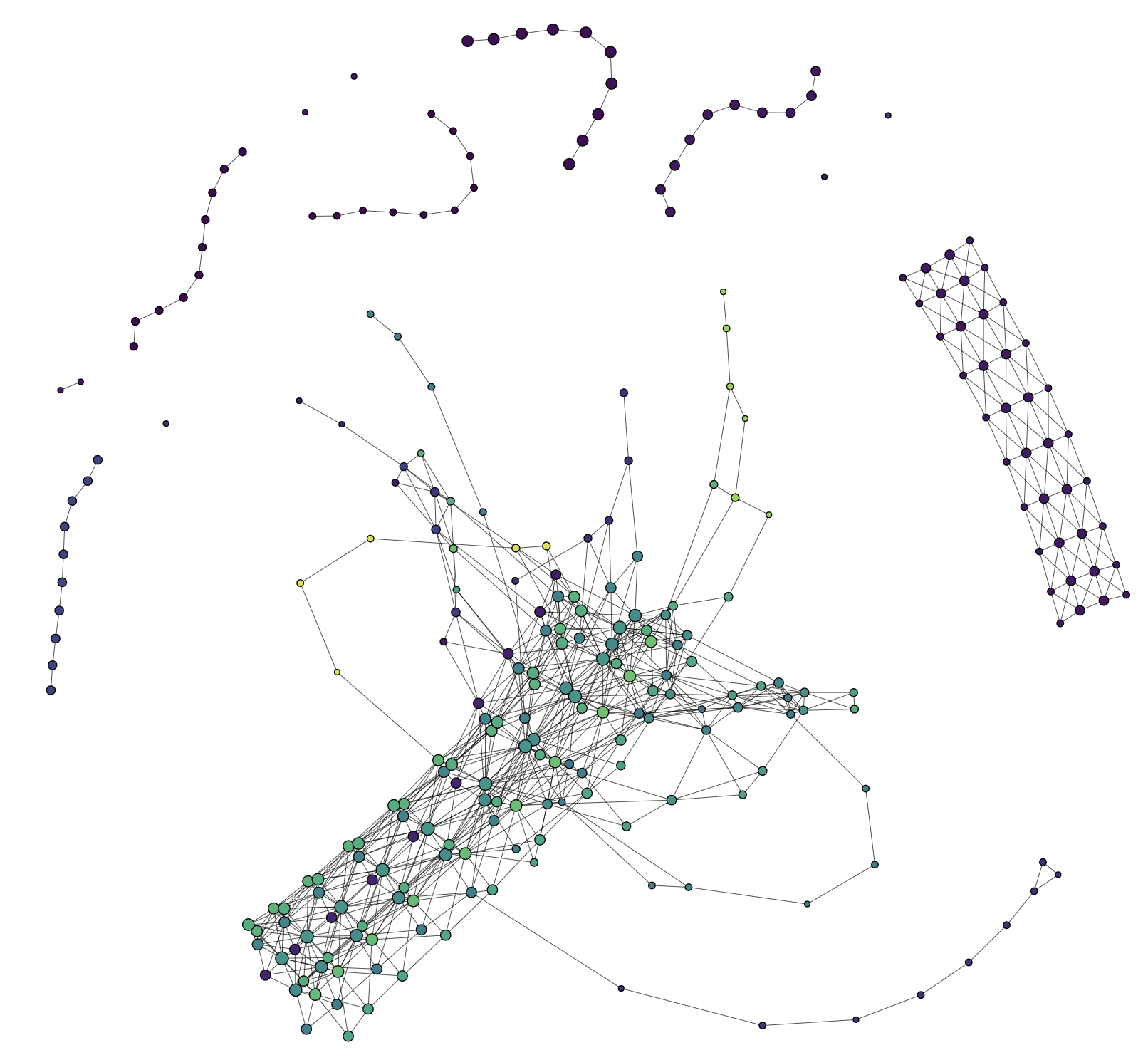}
    \caption{7/3/20 ($n = 10, \delta = 8\%$)}
  \end{subfigure}
  \caption{The evolution of Mapper graphs over time for filtered data (see
  Section~\ref{ssec:filter}).}~\label{fig:four-filtered}
  \end{figure}

For example, Figure~\ref{fig:four-all} shows the evolution of Mapper graphs for
unfiltered data through series of graphs generated from case data reported by 3/6/20,
4/10/20, 5/22/20, and 7/3/20, respectively. Figure~\ref{fig:four-filtered} shows the
evolution of Mapper graphs for filtered data. Since the number of total data points
in each graph varies significantly, we adopt different resolution levels $n$ and
degrees of overlap $\delta$ to preserve relative visual coherence among graphs.

As the number and distribution of U.S. COVID-19 cases evolve over time, the
corresponding Mapper graphs evolve accordingly. New components might break off from
the main trunk as the regions they represent experience aggressive increase in the
number of reported cases. Existing isolated components might also reconnect to the
main trunk through the creation of new edges due to a reduced disparity between the
number of cases reported in the former and its neighboring regions. Additionally,
branches might grow longer as new cases add up.

To illustrate these features, we will in the following sections shift the focus to a
more local level and provide some detail on the evolution of Mapper graphs
representing the local development of the pandemic in two places.

\subsubsection{New York State}
\begin{figure}[h]
  \begin{subfigure}[b]{0.4\textwidth}
    \centering
    \includegraphics[width=\linewidth]{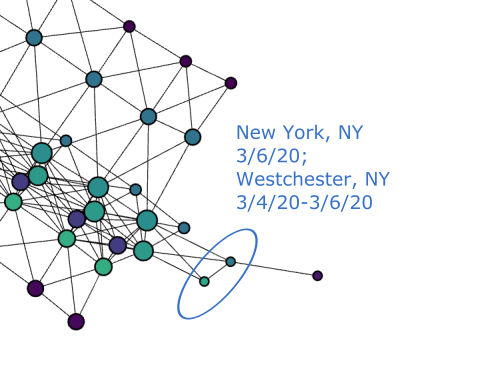}
    \caption{3/6/20 
    }~\label{subfig:evo-ny-a}
  \end{subfigure}
  \begin{subfigure}[b]{0.4\textwidth}
    \centering
    \includegraphics[width=\linewidth]{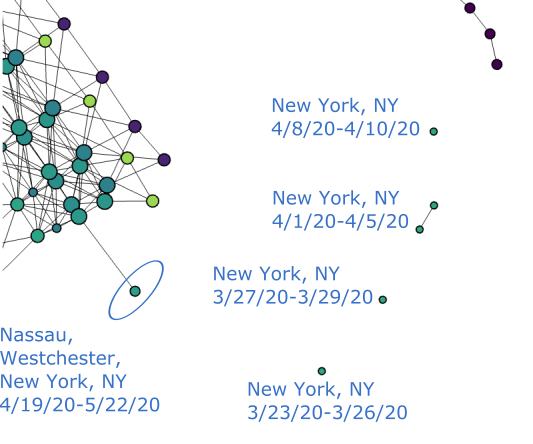}
    \caption{4/10/20 
    }~\label{subfig:evo-ny-b}
  \end{subfigure}
  \begin{subfigure}[b]{0.45\textwidth}
    \centering
    \includegraphics[width=\linewidth]{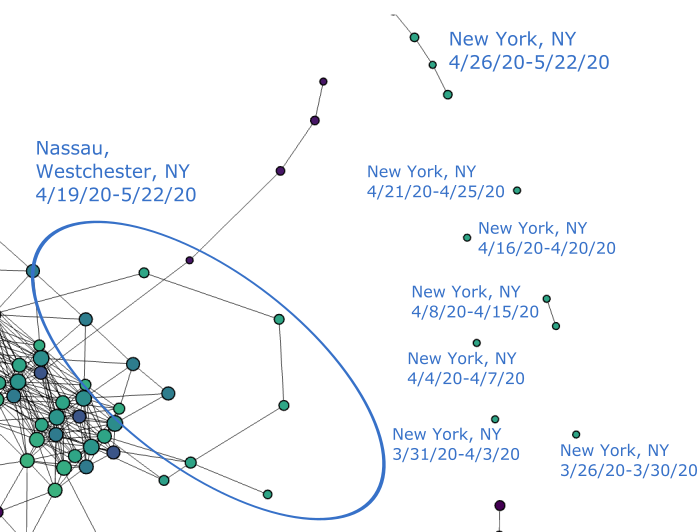}
    \caption{5/22/20 
    }~\label{subfig:evo-ny-c}
  \end{subfigure}
  \begin{subfigure}[b]{0.4\textwidth}
    \centering
    \includegraphics[width=1.1\linewidth]{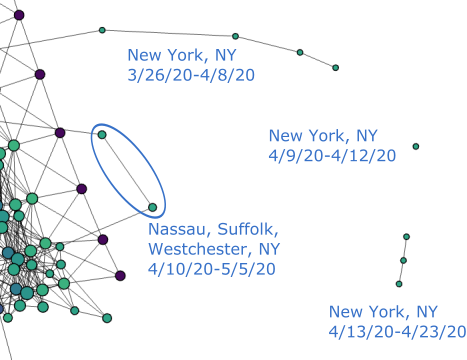}
    \caption{06/19/20 
    }~\label{subfig:evo-ny-d}
  \end{subfigure}
  \begin{subfigure}[b]{0.4\textwidth}
    \centering
    \includegraphics[width=\linewidth]{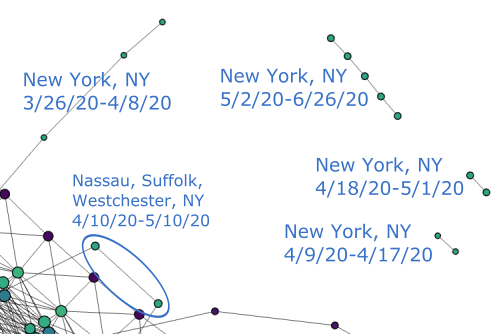}
    \caption{06/26/20 
    }~\label{subfig:evo-ny-e}
  \end{subfigure}
  \begin{subfigure}[b]{0.4\textwidth}
    \centering
    \includegraphics[width=\linewidth]{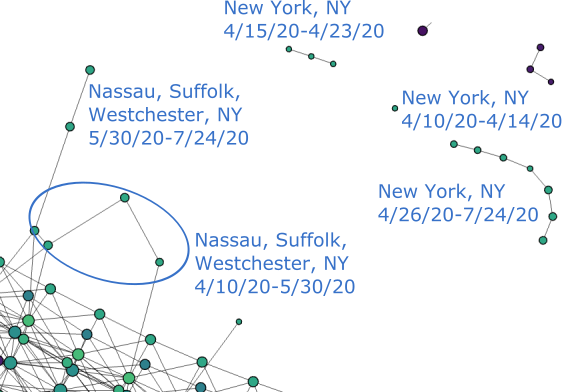}
    \caption{07/24/20 
    }~\label{subfig:evo-ny-f}
  \end{subfigure}
  \caption{The evolution of the New York State cases over time.}~\label{fig:evo-ny}
  \end{figure}

Figure~\ref{fig:evo-ny} gives a few evolutionary snapshots of the branch representing
hot-spots in New York State. In earlier snapshots
(Figures~\ref{fig:evo-ny}(A)--\ref{fig:evo-ny}(D)),
the branch is unstable, with new disconnected nodes emerging frequently through each
update so that the branch appears more fragmented. This reflects the acceleration of
the pandemic in New York during that period. Each update in the data set brings significantly higher case numbers so
that variability of projected data onto the ``number of cases'' dimension causes the
break in the branch.

In ensuing phases, however, the number of cumulative cases is so large that the
relative relevance of each date's numbers gets reduced as a result of the
normalization (see Section~\ref{sssec:normalization}). It thus becomes harder for new
data to affect the shape of the branch. As a result, only the most impactful
developments are picked up and reflected though every update. Hence, when the cases
stabilize in New York, as we see from
Figures~\ref{fig:evo-ny}(D)--\ref{fig:evo-ny}(F), the shape of its branch also tends
to be less volatile. Instead of causing the branch to further break into more
segments as in Figures~\ref{fig:evo-ny}(B) and~\ref{fig:evo-ny}(C), new nodes
continue to build up on the latest segment. This shows that the pandemic no longer
accelerates, notwithstanding a slowed growth. In the end, the structures that remain
in the most recent graph, Figure~\ref{fig:evo-ny}(F), mark the most numerically
relevant turning points in the development of the pandemic in New York, thus
presenting a stable picture documenting the entire process.

In addition, it is also clear that, in contrast to earlier graphs in which New York
is the only prominent branch, we can spot other branching structures in later graphs,
easily discernible in Figure~\ref{fig:four-all}. This displays not only the
development of the pandemic in other areas in parallel with New York, but also the
declining relevance of New York on the national scale. Nevertheless, unlike other
hot-spots whose branch almost disappeared from the graph because of their minimized
relevance (e.g.~most Connecticut hot-spots and King County in Washington), New York's
branch remains highly visible throughout the updates. This is due to the absolute
numerical relevance of New York State's data, which, by 7/23/20, ranks top in the
U.S.~in terms of state-level cumulative COVID-19 cases. Similarly, the branch
representing California and Florida hot-spots also remain visible throughout the
updates because of their absolute national relevance.

\subsubsection{Massachusetts-New Jersey complex}
\begin{figure}[h]
  \begin{subfigure}[b]{0.3\textwidth}
    \centering
    \includegraphics[width=\linewidth]{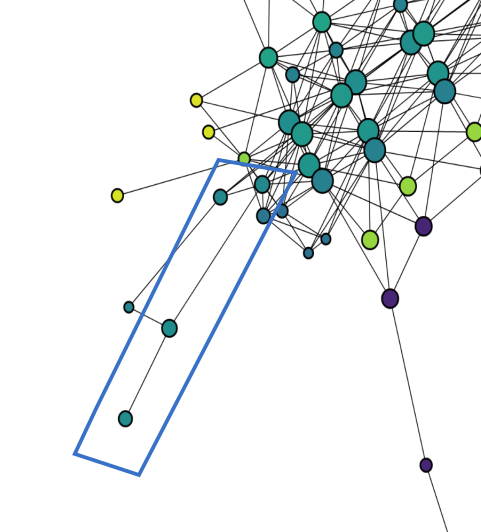}
    \caption{4/10/20 
    }~\label{subfig:evo-manj-a}
  \end{subfigure}
  \begin{subfigure}[b]{0.4\textwidth}
    \centering
    \includegraphics[width=\linewidth]{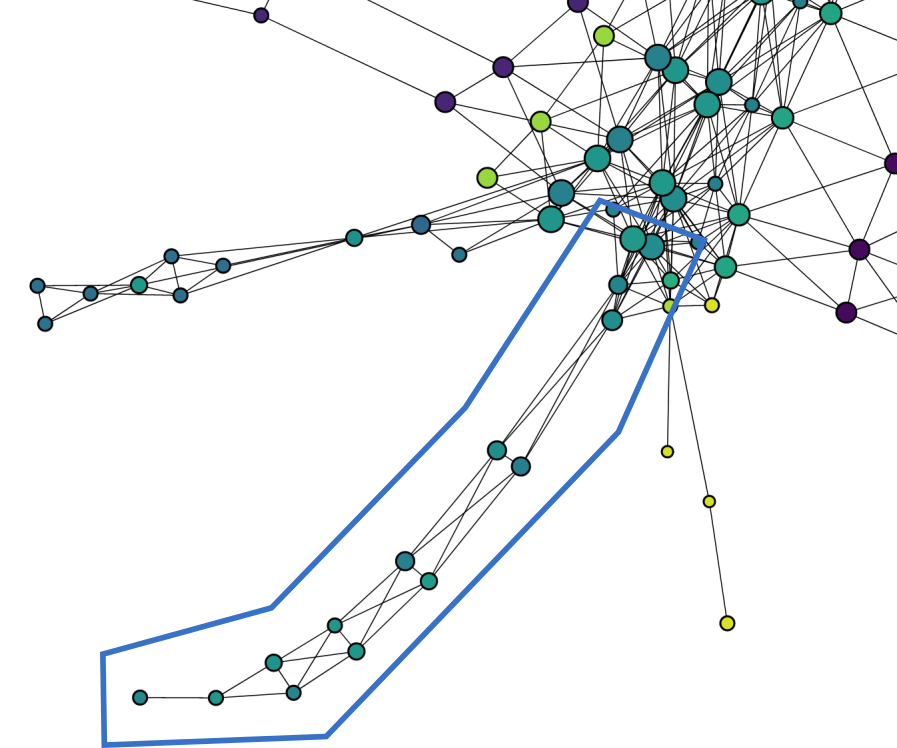}
    \caption{4/24/20 
    }~\label{subfig:evo-manj-b}
  \end{subfigure}
  \begin{subfigure}[b]{0.4\textwidth}
    \centering
    \includegraphics[width=\linewidth]{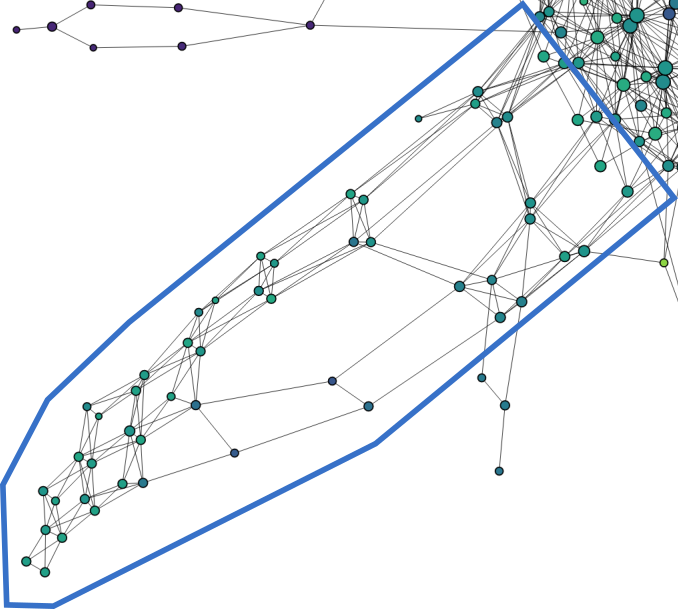}
    \caption{5/22/20 
    }~\label{subfig:evo-manj-c}
  \end{subfigure}
  \begin{subfigure}[b]{0.3\textwidth}
    \centering
    \includegraphics[width=1.1\linewidth]{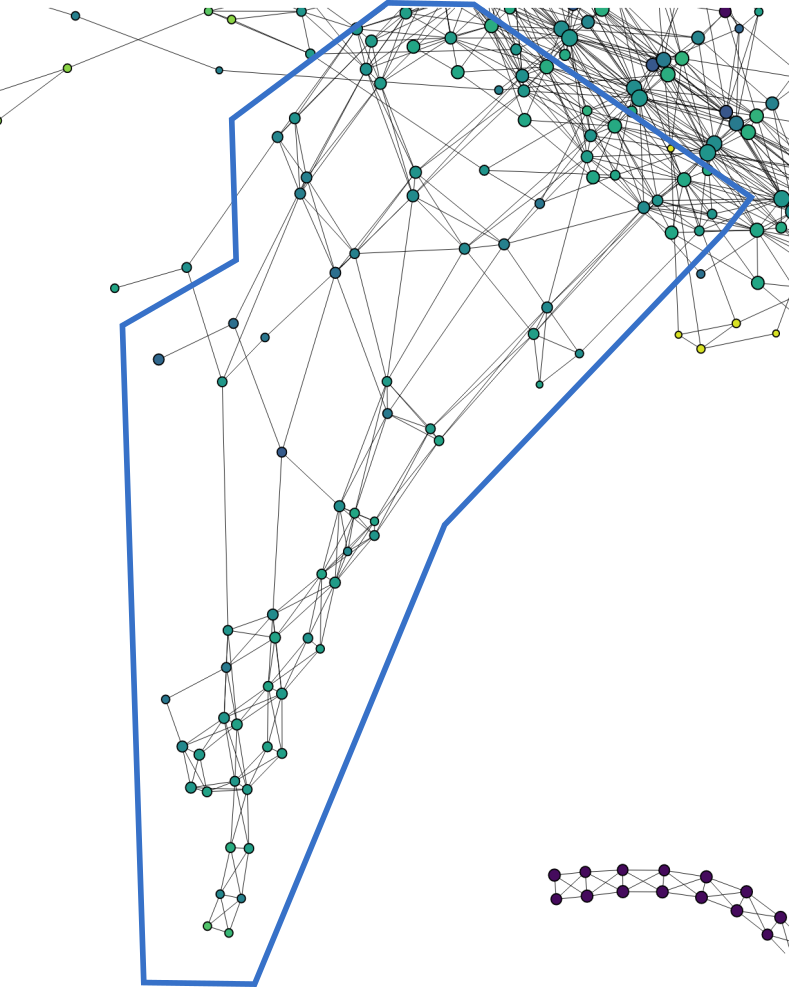}
    \caption{06/19/20 
    }~\label{subfig:evo-manj-d}
  \end{subfigure}
  \begin{subfigure}[b]{0.4\textwidth}
    \centering
    \includegraphics[width=\linewidth]{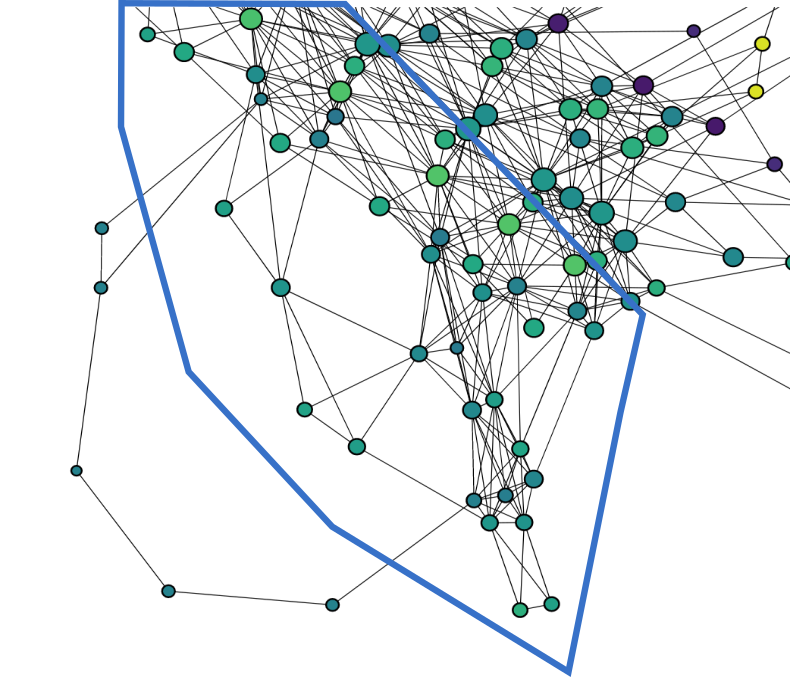}
    \caption{7/3/20 
    }~\label{subfig:evo-manj-e}
  \end{subfigure}
  \begin{subfigure}[b]{0.37\textwidth}
    \centering
    \includegraphics[width=\linewidth]{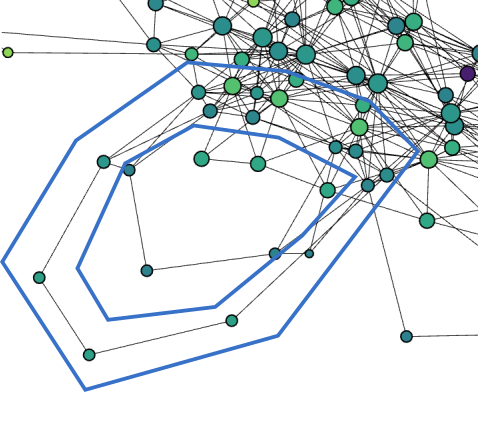}
    \caption{7/24/20 
    }~\label{subfig:evo-manj-f}
  \end{subfigure}
  \caption{The evolution of the Massachusetts-New Jersey complex over
  time.}~\label{fig:evo-manj}
  \end{figure}

As elaborated in Section~\ref{ssec:complex}, because of their geographical
proximity and national prominence, hot-spots in states like Massachusetts, Connecticut
and New Jersey get reflected in the graph in the form of an entangled system of
branches. The evolution of this system can be found in
Figure~\ref{fig:evo-manj}. Similar to New York, these hot-spots have also gone through
a process of acceleration and decline. In contrast, however, the size of this system
is significantly smaller in later snapshots (see Figures~\ref{fig:evo-manj}(E)
and~\ref{fig:evo-manj}(F)). This is because of the reduced national relevance of most
of the hot-spots in this complex, so that their nodes no longer stand out of the main
trunk as they are no longer distinguishable from their neighboring areas. As a
result, only the most severe and most long-lasting hot-spots remained in this
agglomeration, found among more recently developed systems of branches (see
Figure~\ref{fig:evo-manj}(F)).

\section{Conclusions and future work}\label{sec:future}

The present paper provides a case study on the application of the Mapper algorithm to
COVID-19 data collected in the United States. We have shown that Mapper captures a
number of trends in the spread of COVID-19 and provides a more complete picture than
those offered by more standard techniques of data visualization. The existence of
(segmented) branches indicates hot-spots and the emergence of such branches
correlates with troubling trends in the development of the pandemic in the regions
they represent. Geographic proximity between hot-spots is also taken into account, so
that branch complexes arise if nearby counties are similarly experiencing significant
increase in the number of cases. The Mapper algorithm is also capable of tracking the
gradual development of COVID-19 because it incorporates time data, supplying a more
geographically and temporally complete picture of the spread of the virus. Continued
observation on the evolution of the Mapper graphs is therefore clearly desirable.

There are various directions in which future analysis could be employed. For example,
one could replicate the same study for data from different parts of the world. In
addition, one can take into account border closures between countries, for example
within EU or between EU and the rest of Europe.

One could additionally correlate stay-at-home orders issued by each U.S.~state with
critical changes observed in the Mapper graphs in hopes of evaluating the impact and
effectiveness of such orders. Data about travel could also be included in hope that
the Mapper reveals something new about how COVID-19 is spread by travelers through
the U.S. In addition, one could overlay socio-economic data onto the Mapper graphs,
along the lines of what was done in~\cite{DR:COVIDBallMapper}, providing important
insight into how economic factors correlate to the spread of the disease across the
U.S.

Unfortunately, the visualizations that we so far produced through the Mapper
algorithm cannot offer strict and accurate predictions for the future development of
the pandemic, since such predictions require the consideration of more real-life
factors than we studied in this paper. But those factors can be incorporated by
imposing more structure onto the Mapper. For example, we could define branches more
rigorously using graph-theoretic notions. The length of the branches and the degree
of the nodes in them can be assigned real-life meaning, along the lines of what was
done by Escolar et.~al.~in~\cite{EHIO:MapperFirms}. This would provide more insight
and understanding into the development of hot-spots, and could lead to a more
predictive model given by the Mapper.

Another direction is to apply a different popular version of topological data
analysis called \emph{persistent homology}. In this setup, the data is turned into a
topological space and one then studies the space's connected components as well as
``holes'' of various dimensions. Persistent homology has been used to great effect in
a number of settings, including biology, neuroscience, medicine, materials science,
sensor networks, financial networks, and many others. For introductions to persistent
homology and more details about its applications, see~\cite{Carlsson:TDA,
Carlsson:TDAHomotopy, EH:CompTop, OPTGH:TDAOverview}. The usefulness of persistent
homology as a predictive epidemiological tool was demonstrated by Lo and
Park~\cite{LP:TDAZika} in their study of the Zika virus; emulating this approach
might also lead to the construction of a useful predictive model for the coronavirus.



\clearpage

\bibliographystyle{amsplain}

\end{document}